\documentclass[12pt]{article}
\usepackage[pdftex]{graphicx}
\usepackage{epsfig,cite}
\usepackage {amsmath}
\usepackage{amssymb}
\usepackage{slashed}
\newcount\timecount
\newcount\hours \newcount\minutes  \newcount\temp \newcount\pmhours
\hours = \time
\divide\hours by 60
\temp = \hours
\multiply\temp by 60
\minutes = \time
\advance\minutes by -\temp
\def\hour{\the\hours}
\def\minute{\ifnum\minutes<10 0\the\minutes
            \else\the\minutes\fi}
\def\clock{
\ifnum\hours=0 12:\minute\ AM
\else\ifnum\hours<12 \hour:\minute\ AM
      \else\ifnum\hours=12 12:\minute\ PM
            \else\ifnum\hours>12
                 \pmhours=\hours
                 \advance\pmhours by -12
                 \the\pmhours:\minute\ PM
                 \fi
            \fi
      \fi
\fi
}

\def\monthname{\relax\ifcase\month 0/\or January\or February\or
   March\or April\or May\or June\or July\or August\or September\or
   October\or November\or December\else\number\month/\fi}

\def\bold#1{\setbox0=\hbox{$#1$}%
     \kern-.025em\copy0\kern-\wd0
     \kern.05em\copy0\kern-\wd0
     \kern-.025em\raise.0433em\box0 }

\def\beq{\begin{equation}}
\def\eeq{\end{equation}}

\def\ga{\mathrel{\raise.3ex\hbox{$>$\kern-.75em\lower1ex\hbox{$\sim$}}}}
\def\la{\mathrel{\raise.3ex\hbox{$<$\kern-.75em\lower1ex\hbox{$\sim$}}}}
\def\gev{{\rm \, Ge\kern-0.125em V}}
\def\tev{{\rm \, Te\kern-0.125em V}}
\def\gyr{{\rm \, G\kern-0.125em yr}}

\def\gappeq{\mathrel{\rlap {\raise.5ex\hbox{$>$}}
{\lower.5ex\hbox{$\sim$}}}}
\def\lappeq{\mathrel{\rlap{\raise.5ex\hbox{$<$}}
{\lower.5ex\hbox{$\sim$}}}}
\def\Toprel#1\over#2{\mathrel{\mathop{#2}\limits^{#1}}}

\def\m12{m_{1\!/2}}

\def\bea{\begin{eqnarray}}
\def\eea{\end{eqnarray}}

\def\beqar{\begin{eqnarray}}
\def\eeqar{\end{eqnarray}}

\usepackage{lscape}

\def\beq{\begin{equation}}
\def\eeq{\end{equation}}

\begin{document}

\begin{titlepage}
\pagestyle{empty}
\rightline{KCL-PH-TH/2012-47, LCTS/2012-33, CERN-PH-TH/2012-343}
\rightline{UMN--TH--3128/12, FTPI--MINN--12/39}
\vspace{0.5cm}
\begin{center}
{\large {\bf The Higgs Mass beyond the CMSSM }}

\end{center}
\vspace{0.5cm}
\begin{center}
{\bf John~Ellis}$^{1,2}$,
{\bf Feng Luo}$^{1}$, {\bf Keith~A.~Olive}$^{3}$
and {\bf Pearl Sandick}$^{4}$\\
\vskip 0.2in
{\small {\it
$^1${Theoretical Physics and Cosmology Group, Department of Physics, King's College London, London~WC2R 2LS, UK}\\
$^2${TH Division, Physics Department, CERN, CH-1211 Geneva 23, Switzerland}\\
$^3${William I. Fine Theoretical Physics Institute, School of Physics and Astronomy,\\
University of Minnesota, Minneapolis, MN 55455,\,USA}\\
$^4${Department of Physics, University of Utah, Salt Lake City, Utah 84112, USA}} \\
}
\vspace{1cm}
{\bf Abstract}
\end{center}
{\small
The apparent discovery of a Higgs boson with mass $\sim 125$~GeV has had 
a significant impact on the constrained minimal supersymmetric extension of the Standard Model
in which the scalar masses, gaugino masses and tri-linear $A$-terms are assumed to be universal at the GUT scale (the CMSSM).  
Much of the low-mass parameter space in the CMSSM has been excluded by supersymmetric particle searches at the LHC
as well as by the Higgs mass measurement and the emergent signal for $B_s \to \mu^+ \mu^-$.
Here, we consider the impact of these recent LHC results on several variants of the CMSSM
with a primary focus on obtaining a Higgs mass of $\sim 125$~GeV. 
In particular, we consider the one- and two-parameter extensions of the CMSSM
with one or both of the Higgs masses set independently of the common
sfermion mass, $m_0$ (the NUHM1,2). We also consider the one-parameter extension
of the CMSSM in which the input universality scale $M_{in}$ is below the GUT scale (the sub-GUT CMSSM).
We find that when $M_{in} < M_{GUT}$ large regions of parameter space open up
where the relic density of neutralinos can successfully account for dark matter
with a Higgs boson mass $\sim 125$~GeV. In some of these regions essential roles are 
played by coannihilation processes that are usually less important in the CMSSM with 
$M_{in} = M_{GUT}$. Finally, we reconsider mSUGRA models with
sub-GUT universality, which have the same number of parameters as the CMSSM.
Here too, we find phenomenologically viable regions of parameter space, which
are essentially non-existent in GUT-scale mSUGRA models.  Interestingly, we find that
the preferred range of the $A$-term straddles that predicted by the simplest Polonyi model. }


\vfill
\leftline{December 2012}
\end{titlepage}

\section{Introduction}

Supersymmetry is one of the best-motivated extensions of the Standard Model,
including the naturalness of the electroweak scale, the appearance of a candidate
for the cold dark matter if $R$-parity is conserved, unification of the fundamental
interactions, the prediction of a Higgs boson weighing $\lappeq 130$~GeV, and the
stabilization of the electroweak vacuum if the Higgs boson weighs $\lappeq127$~GeV,
as seems now to be the case. For these and other, more theoretical reasons, 
much attention has been focused on models with supersymmetric particles weighing
$\lappeq 1$~TeV, and notably the simplest, minimal supersymmetric extension of the
Standard Model (the MSSM). In particular, many studies have been made of the
simplified variant in which the soft supersymmetry-breaking scalar masses, gaugino 
masses and tri-linear $A$-terms are assumed to originate at some high mass scale
and be universal at the supersymmetric GUT scale (the CMSSM) \cite{funnel,cmssm,efgosi,cmssm2,cmssmwmap}.

However, there is as yet no direct evidence for supersymmetry. Some years ago,
hopes were high that the deviation of the experimental measurement of the anomalous
magnetic moment of the muon, $g_\mu - 2$~\cite{newBNL}, from the prediction of the Standard Model
might herald the appearance of low-mass supersymmetric particles accessible to the
initial run of the LHC. However, this has not yet transpired, and the continuing absence
of supersymmetric particles at the LHC~\cite{lhc} is putting increasing
pressure on the CMSSM and rendering the supersymmetric interpretation of $g_\mu - 2$
somewhat dubious in this and related models.

The ATLAS and CMS experiments have recently discovered a new boson
with spin $\ne 1$ and couplings similar to those of the Standard Model Higgs boson \cite{lhch}. This is
compatible with such simple supersymmetric models, which predict not only the existence
of a Higgs boson with mass $\lappeq 130$~GeV, but also that its couplings to Standard
Model particles should be indistinguishable from those in the Standard Model at the present
level of accuracy. Using {\tt FeynHiggs} \cite{FH} there is an uncertainty $\sim \pm 1.5$~GeV in the prediction for the Higgs 
mass for any given set of input CMSSM parameters, but the measurement of a mass
$\sim 125$ to $126$~GeV does disfavour relatively low sparticle masses. Taken together
with the absence so far of supersymmetric particles at the LHC, the Higgs mass measurement
excludes a large range of low-mass CMSSM parameters~\cite{mc8,postlhc,mc75,125-other,eo6}.

In parallel, increasing pressure is being placed
on models with large $\tan \beta$ by the strengthening experimental constraint on the decay 
$B_s \to \mu^+ \mu^-$~\cite{bmm}, in particular the recent detection by LHCb
of $B_s \to \mu^+ \mu^-$ with a value close to the Standard Model
prediction~\cite{:2012ct}. Currently, the 95\% upper limit on the branching fraction relative to the Standard Model
value is 1.65 from LHCb alone, improving to 1.50 when combined with other experiments.

These developments motivate the study of generalizations of the CMSSM, the search for viable 
alternatives, and the exploration of their observational signatures, with a view to future
searches at the LHC and elsewhere.

In this spirit, here we consider various one- and two-parameter extensions of the CMSSM.
One option that has often been studied is to allow one or both of the soft supersymmetry-breaking
contributions to Higgs scalar masses to be independent of the common sfermion mass, $m_0$ 
(the NUHM1,2)~\cite{nuhm1,eosknuhm,nonu,nuhm2}. 
These offer some extra possibilities for accommodating the
mass of the Higgs boson and the absence (so far) of supersymmetric particles at the LHC while
yielding the appropriate relic cold dark matter density. 

Another way to extend
the CMSSM with one additional parameter is to allow the input universality scale to be lower 
than the GUT scale~\cite{subGUT} (the sub-GUT CMSSM). 
In this case, the sparticle spectrum is compressed relative to that
in the CMSSM, because of the reduced amount of renormalization of the soft
supersymmetry-breaking masses if they are universal below the GUT scale. This in turn implies
that relatively large ranges of sparticle masses are compatible simultaneously
with a Higgs boson mass $\sim 125$~GeV and an appropriate relic dark matter
density, often involving novel coannihilation mechanisms not relevant in the CMSSM. 

As a corollary, one may impose an additional constraint on the sub-GUT extension of the
CMSSM and retain many of its phenomenological advantages. In particular, one may impose
the specific relation $A_0 = B_0 + m_0$ between the trilinear and bilinear soft supersymmetry-breaking
parameters $A_0$ and $B_0$ that is characteristic of 
minimal supergravity (mSUGRA) theories~\cite{Fetal,acn,bfs}.
Intriguingly, we find acceptable models for a range of $A_0$ values that includes the specific
value $A_0 = (3 - \sqrt{3}) \times m_{3/2}$ found in the original Polonyi model \cite{Polonyi}.

A generic feature of the NUHM1,2 and such sub-GUT models is the increased importance of
coannihilation effects, that are only important along narrow strips in the CMSSM parameter space.
Accordingly, in this paper we have paid particular attention to the treatment of coannihilation processes, using
a more complete treatment than in our previous analyses of the 
NUHM1,2~\cite{eosknuhm,nuhm2} and sub-GUT models~\cite{subGUT}.

The layout of this paper is as follows. In Section~2 we recall relevant features of gravity-mediated 
models such as mSUGRA and the CMSSM, as well as
pointing out some important features of 
their extensions to the NUHM and sub-GUT 
models. In Section 3, we discuss in more detail the impact of the LHC constraints and
briefly review the current constraints on the CMSSM and mSUGRA. Section~4 discusses the impact
of the LHC constraints on the NUHM1,2, Section~5 discusses sub-GUT generalizations
of the CMSSM, and Section~6 discusses the specific case of sub-GUT mSUGRA. Finally, 
Section~7 summarizes our principal conclusions.

\section{mSUGRA, the CMSSM and Generalizations}

\subsection{mSUGRA}

We start by recalling relevant aspects of minimal supergravity  (mSUGRA) models, which have
a flat K\"ahler potential and a low-energy scalar potential that can be written as~\cite{Fetal,acn,bfs}
 \begin{eqnarray}
V  & =  &  \left|{\partial W \over \partial \phi^i}\right|^2 +
\left( A_0 W^{(3)} + B_0 W^{(2)} + h.c.\right)  + m_{3/2}^2 \phi^i \phi_i^*  \, ,
\label{pot}
\end{eqnarray}
where $W$ is the superpotential for matter fields, 
\beq
W =  \bigl( y_e H_1 L e^c + y_d H_1 Q d^c + y_u H_2
Q u^c \bigr) +  \mu H_1 H_2  \, ,
\label{WMSSM}
\eeq 
and the SU(2) indices have been suppressed. In (\ref{pot}), $W^{(3)}$ is the trilinear part of the superpotential,
$W^{(2)}$ is the bilinear part, and $m_{3/2}$ is the gravitino mass. This is the order parameter for local
supersymmetry breaking, and in mSUGRA one finds scalar mass universality, $m_0 = m_{3/2}$.
This condition is applicable at some input renormalization scale, $M_{in}$, which is usually associated
with the grand unification scale. In addition, in mSUGRA there is a relation
between the tri- and bilinear supersymmetry breaking terms, $A_0 = B_0 + m_0$.
With a minimal choice for the gauge kinetic function, gaugino mass universality characterized by $m_{1/2}$ is also
obtained, which is usually imposed at the same input scale $M_{in}$.

Minimization of the Higgs potential leads to two vacuum conditions at the weak scale, which 
can be expressed as
\beq
\mu^2=\frac{m_1^2-m_2^2\tan^2\beta+\frac{1}{2}m_Z^2(1-\tan^2\beta)+\Delta_{\mu}^{(1)}}{\tan^2\beta-1+\Delta_{\mu}^{(2)}} \, ,
\label{eq:mu}
\eeq
and 
\beq
B \mu = - \frac{1}{2}(m_1^2+m_2^2+2\mu^2)\sin 2\beta +\Delta_B \, ,
\label{eq:muB}
\eeq  
where $m_{1,2}$ are the soft supersymmetry-breaking Higgs masses (evaluated at the weak scale), 
$\tan \beta$ is the ratio of the two Higgs vacuum expectation values, and $\Delta_B$ and $\Delta_\mu^{(1,2)}$ are loop
corrections~\cite{Barger:1993gh}.
As a result, an mSUGRA model can be defined in terms of just 3 continuous parameters:
$m_{1/2}$, $m_0$ and $A_0$, and the sign of $\mu$. Together, these determine $\tan \beta$
and the magnitude of $\mu$ via the 
vacuum conditions (\ref{eq:mu}, \ref{eq:muB}). 

\subsection{The CMSSM}

The CMSSM is the most thoroughly studied of 
all classes of constrained supersymmetric models~\cite{funnel,cmssm,efgosi,cmssm2,cmssmwmap}.
It is  a two-parameter generalization of mSUGRA, in which
the relation between $A_0$ and $B_0$ is dropped, allowing
one to treat $\tan \beta$ as a free parameter,  as well as dropping the
relationship between the universal sfermion mass $m_0$ and the gravitino mass, which is possible if the
supergravity K\"ahler potential includes non-minimal kinetic terms.
Thus, the CMSSM is defined by the sign of $\mu$, $m_{1/2}$, $m_0$, $A_0$, $\tan \beta$, and $m_{3/2}$,
though the latter is usually ignored, e.g., because it is assumed implicitly to be irrelevantly large. 
We note that the CMSSM can be directly related to mSUGRA~\cite{dmmo}
through an extension in which terms proportional to $W^{(2)}$ are added to the K\"ahler
potential as in the Giudice-Masiero mechanism~\cite{gm}. 
For example, a contribution to $K$ of the form
\beq
\Delta K = c_H H_1 H_2  + h.c. \, ,
\label{gmk}
\eeq
where $c_H$ is a constant, and $H_{1,2}$ are the usual MSSM
Higgs doublets,
affects  the boundary conditions for 
both $\mu$ and the $B$ term at the supersymmetry breaking input scale, $M_{in}$,
in a manner similar to the CMSSM.

\subsection{The NUHM1,2}

One of the possible one-parameter extensions of the CMSSM is
the oft-studied NUHM1~\cite{nuhm1,eosknuhm}, in which
the soft supersymmetry-breaking Higgs masses are taken to be equal to each other, but are allowed to differ
from the otherwise universal sfermion mass, $m_1 = m_2 \ne m_0$. 
We also discuss here a two-parameter extension of the CMSSM, namely
the NUHM2~\cite{nuhm2} in which $m_1 \ne m_2 \ne m_0$, in general.

\subsection{Sub-GUT versions of CMSSM and mSUGRA}

Another one-parameter generalization of the CMSSM is provided by models 
in which the input scale for supersymmetry breaking universality differs
from the GUT scale. Although $M_{in}$ might be chosen above the GUT scale~\cite{superGUT,dmmo,dlmmo}, 
such a choice would introduce many more parameters associated with different GUT models, and
its study lies beyond the scope of this work.
Here we concentrate on `sub-GUT' models in which $M_{in} < M_{GUT}$~\cite{subGUT},
which can be related to mirage unification models~\cite{mixed}.

Both the NUHM1 and sub-GUT CMSSM are six-parameter models (five if one neglects the gravitino mass,
e.g., by assuming that it is irrelevantly large).
As we shall see, both the NUHM1,2 and sub-GUT models allow a suitable relic dark matter
density and a Higgs mass of 125 GeV to be achieved simultaneously. This is also the case in sub-GUT versions of 
mSUGRA. This is a four-parameter class of models ($m_{1/2}, m_{3/2}, A_0$, and $M_{in}$).
Interestingly, we find that such models are phenomenologically viable in a relatively
restricted range for $A_0$, which straddles the Polonyi value $A_0 = (3 -\sqrt{3}) \times
m_{3/2}$~\cite{Polonyi,bfs}.

\begin{figure}[htb!]
\begin{minipage}{8in}
\includegraphics[height=3.5in]{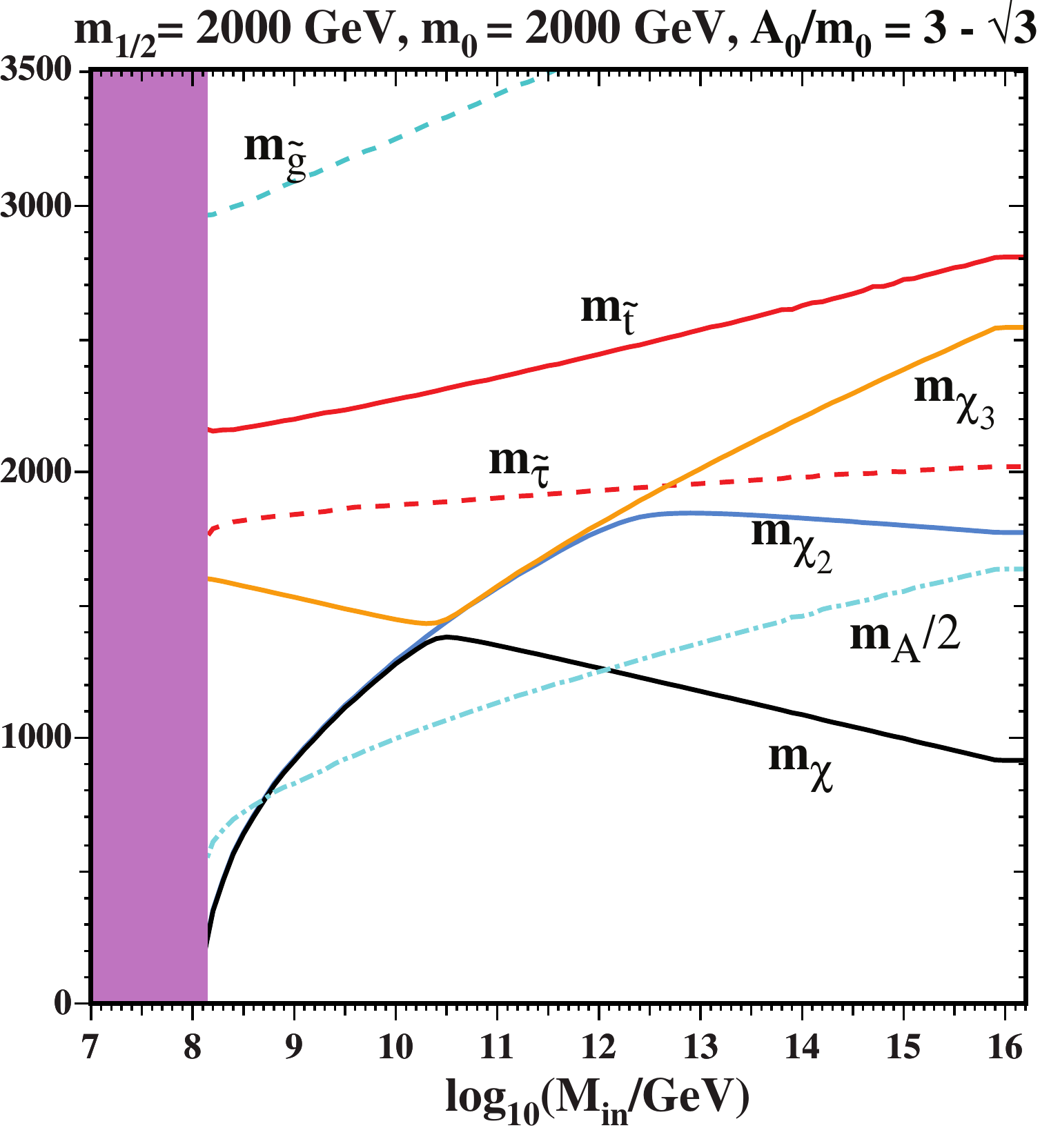}
\hspace*{0.17in}
\includegraphics[height=3.5in]{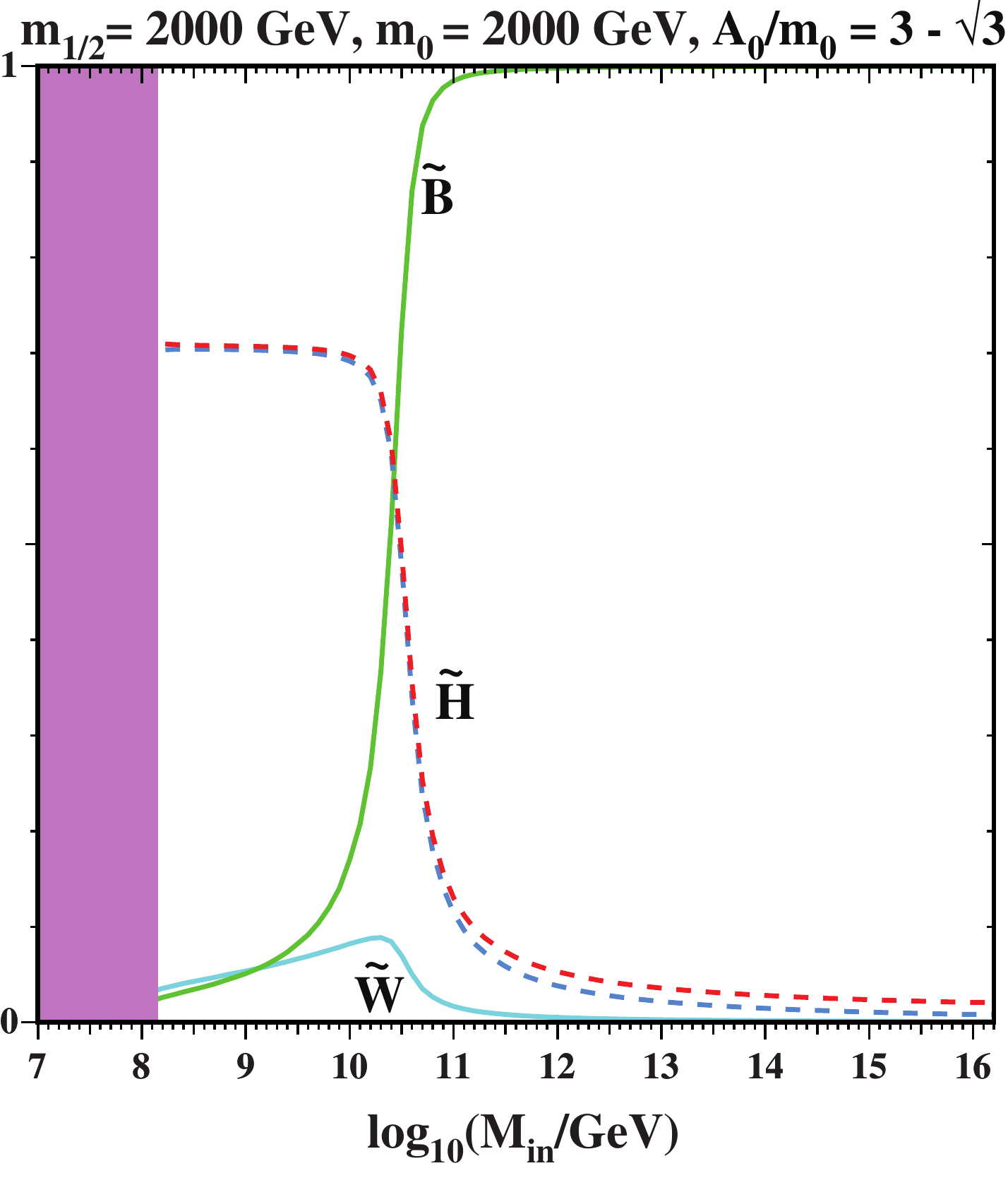}
\hfill
\end{minipage}
\caption{
{\it
The evolution of (left) the sparticle spectrum and (right) the composition of the
LSP $\chi$ as functions of $M_{in}$ in a specific sub-GUT Polonyi mSUGRA
scenario with $m_{1/2} = m_0 = 2000$ GeV.
}} 
\label{fig:subGUTmSUGRA} 
\end{figure}

As an appetizer for the subsequent discussion, we display in the left panel of Fig.~\ref{fig:subGUTmSUGRA} the
evolution of the sparticle spectrum with $M_{in} < M_{GUT}$ for a typical Polonyi mSUGRA
scenario with $m_{1/2} = m_0 = 2000$ GeV. At these values of $(m_{1/2}, m_0)$, $\tan \beta$ varies
from $\sim 18$ when  $M_{in} = M_{GUT}$ to $\sim 43$ when $M_{in} = 3 \times 10^8$ GeV:
we do not find consistent electroweak vacuum solutions much below this value of $M_{in}$,
as indicated by the mauve shading in Fig.~\ref{fig:subGUTmSUGRA}.
As $M_{in}$ decreases, we see that the spectrum compresses until $M_{in} \sim 10^{10.5}$~GeV
with, in particular, the approach of $m_{\chi_3}$ and $m_{\chi_2}$ to $m_{\chi}$.
This has the general effect of enhancing the importance of coannihilation processes that are
less suppressed by Boltzmann factors $\sim {\rm exp}(- \Delta m/T)$. 
We also note that $m_\chi \simeq m_A/2$ when $M_{in} \sim 10^{12}$~GeV, indicating
that the parameters traverse a rapid s-channel annihilation funnel region.
There is level crossing in the
neutralino spectrum when $M_{in} \sim 10^{10.5}$~GeV, with the composition of the lightest supersymmetric particle (LSP) $\chi$
making a transition from almost pure bino to almost pure Higgsino, as seen in the right panel
of Fig.~\ref{fig:subGUTmSUGRA}. For $M_{in} \lappeq 10^{10.5}$~GeV, the LSP and the next-to-lightest supersymmetric particle 
(NLSP) are nearly degenerate Higgsinos, with
important implications for annihilation rates. We also note that
their masses approach $m_A/2$ again just below $10^9$ GeV, 
so that rapid (co)annihilations via s-channel $H/A$ poles again become important. 

This example illustrates just some of the variety of (co)annihilation
phenomena that can control the relic LSP density in sub-GUT models.
More of the possible richness will be seen in subsequent figures.

\section{The Impacts of Recent LHC Results on the CMSSM and mSUGRA}

The following are the principal experimental, phenomenological and cosmological constraints
on supersymmetric models that we consider in this paper:

\begin{itemize}

\item The preferred cosmological cold dark matter density range is~\cite{wmap}: $\Omega_{CDM} h^2 = 0.112 \pm 0.006$, where $h$ is the present Hubble expansion rate in units of 100 km/s/Mpc. We
emphasize models in which the lightest supersymmetric particle (LSP), assumed here to be
the lightest neutralino, $\chi$, is the dominant source of cold dark matter~\cite{ehnos}. In general, supersymmetric models
in which the LSP density is not excessive are squeezed closer to the boundaries where the LSP becomes
charged or there is no consistent solution to the electroweak vacuum conditions (\ref{eq:mu}, \ref{eq:muB}),
and do not have very different values of the input parameters. 

\item The strongest upper limit on the spin-independent cold dark matter scattering cross section
is from XENON100 \cite{XENON100}, which excludes the
focus-point region of the CMSSM and some analogous regions in models with more parameters
\cite{mc8}.

\item The anomalous magnetic moment of the muon, $g_\mu - 2$, will be treated as an optional
constraint \cite{newBNL}, unlike the previous constraints, which are considered to be mandatory.

\item Constraints are provided by flavour physics, in particular by $b \to s \gamma$~\cite{bsgex} 
and $B_s \to \mu^+ \mu^-$~\cite{bmm,:2012ct},
the latter now being dominated by the measurement by LHCb and upper limits from the other LHC experiments.

\item ATLAS and CMS provide limits from
searches for missing-energy events at the LHC, in particular the ATLAS search with $\sim 5$/fb of
data at 8~TeV \cite{lhc,ATLAS}, which is currently the most sensitive public result.

\item The measured mass of the (lightest) Higgs boson is taken as $M_h = 125.7 \pm 1.0$~GeV \cite{lhch}.
We recall that the theoretical calculation uncertainty of $M_h$ for any given set of input parameters for the CMSSM
or a similar model is typically $\sim 1.5$~GeV, so any parameter set yielding $M_h \ge 124$~GeV
could probably be regarded as acceptable.

\end{itemize}

\subsection{Impacts on CMSSM models}

\begin{figure}[htb!]
\begin{minipage}{8in}
\includegraphics[height=3.3in]{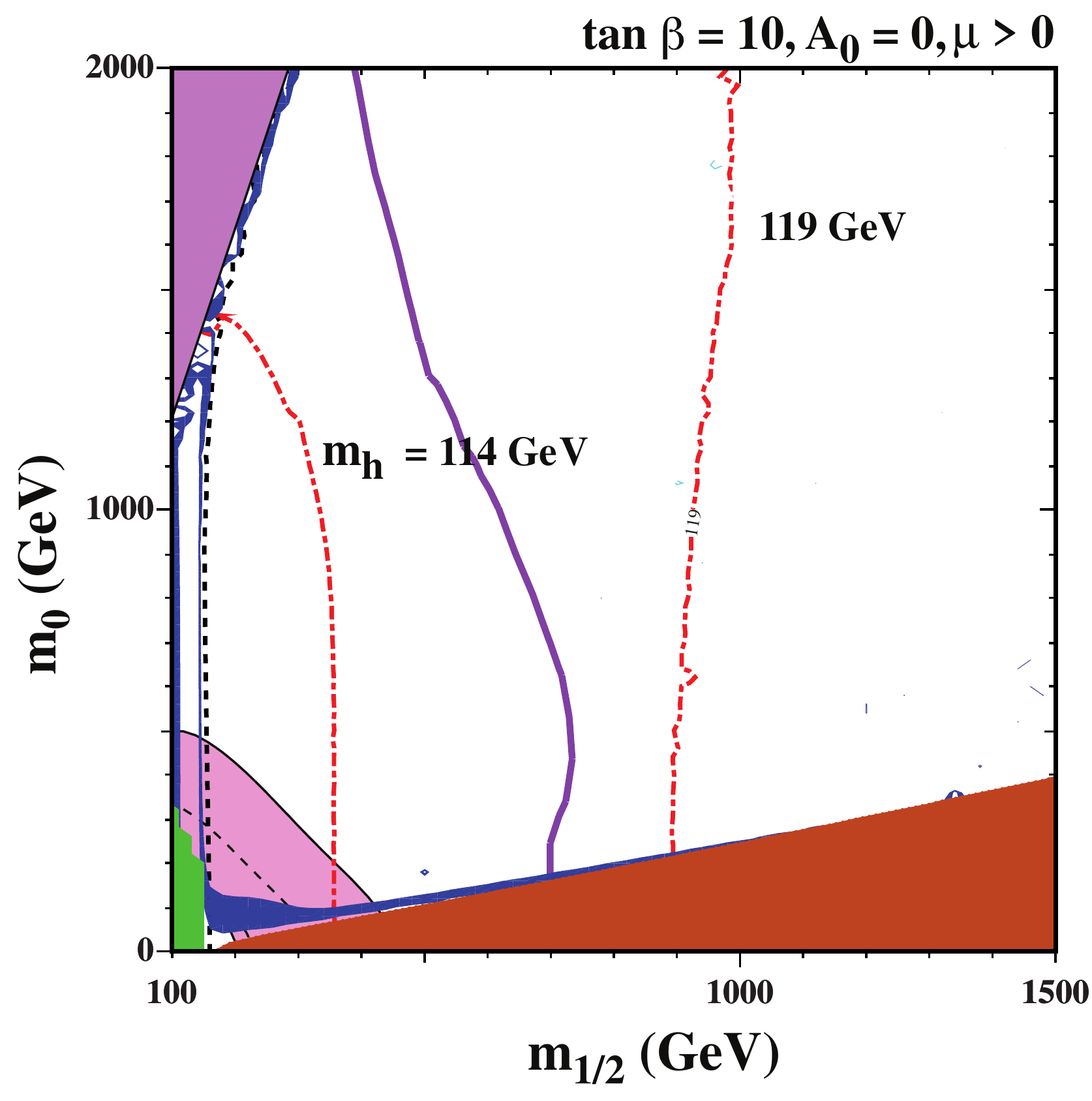}
\hspace*{-0.17in}
\includegraphics[height=3.3in]{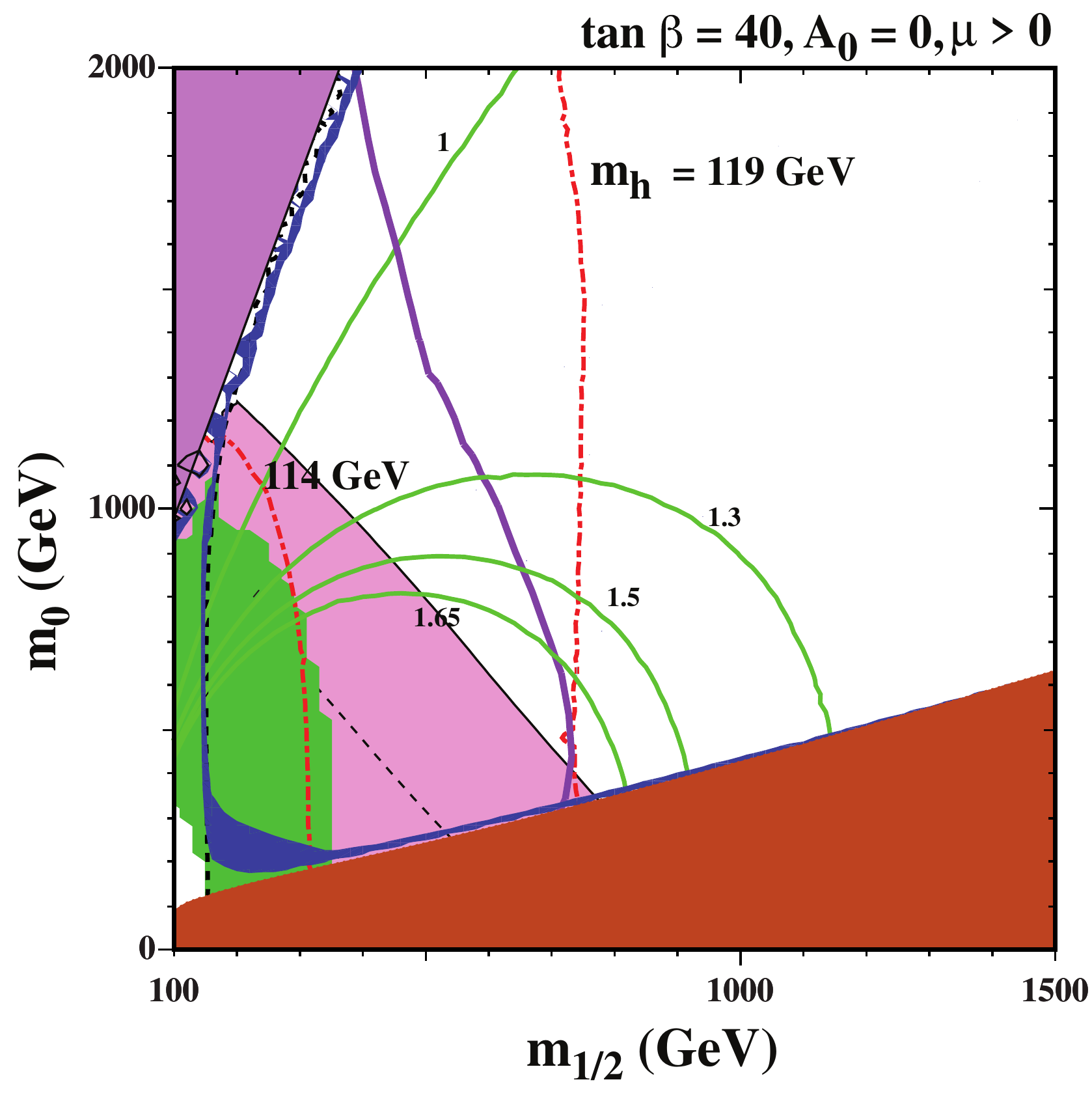}
\hfill
\end{minipage}
\begin{minipage}{8in}
\includegraphics[height=3.3in]{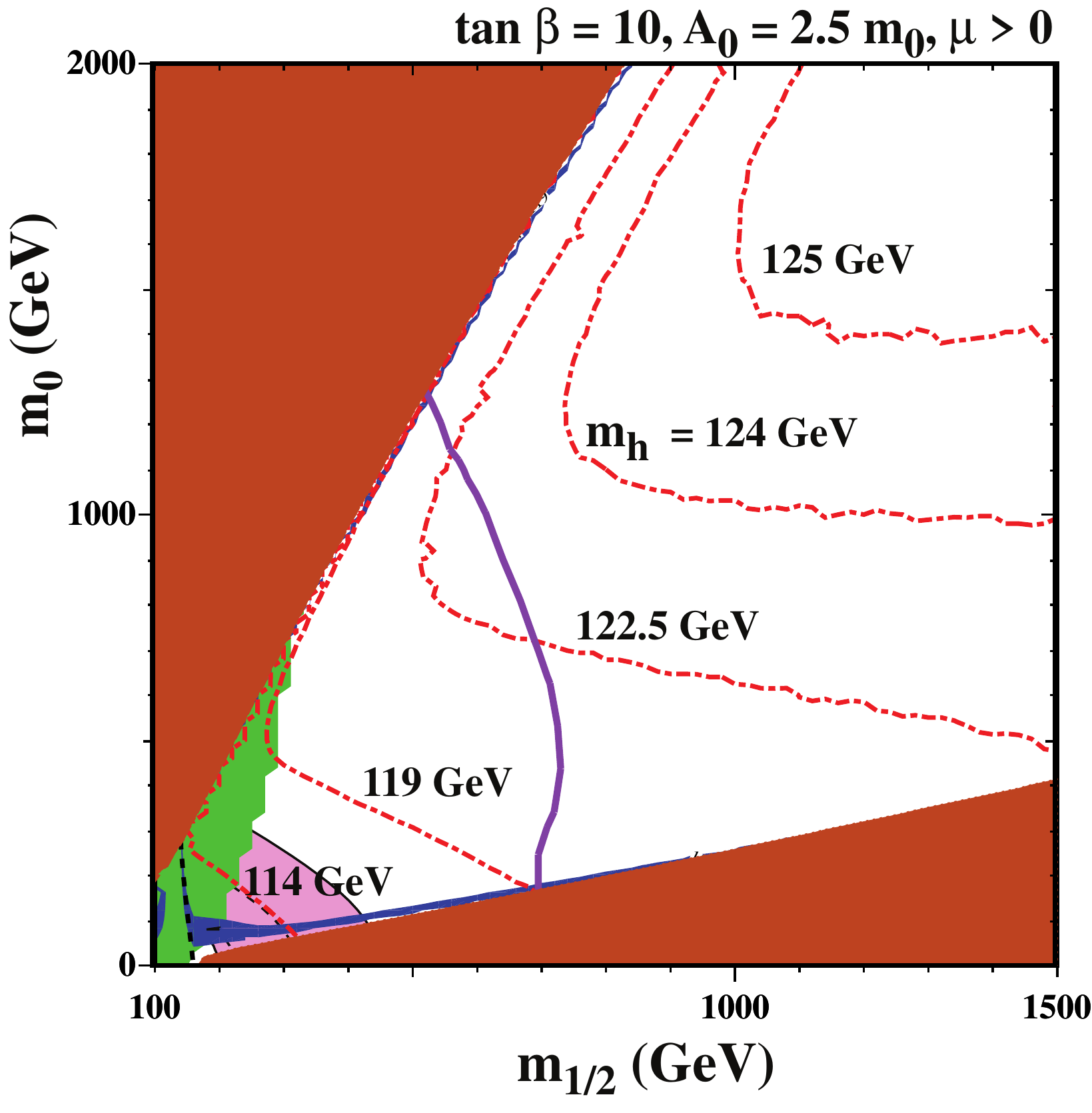}
\hspace*{-0.17in}
\includegraphics[height=3.3in]{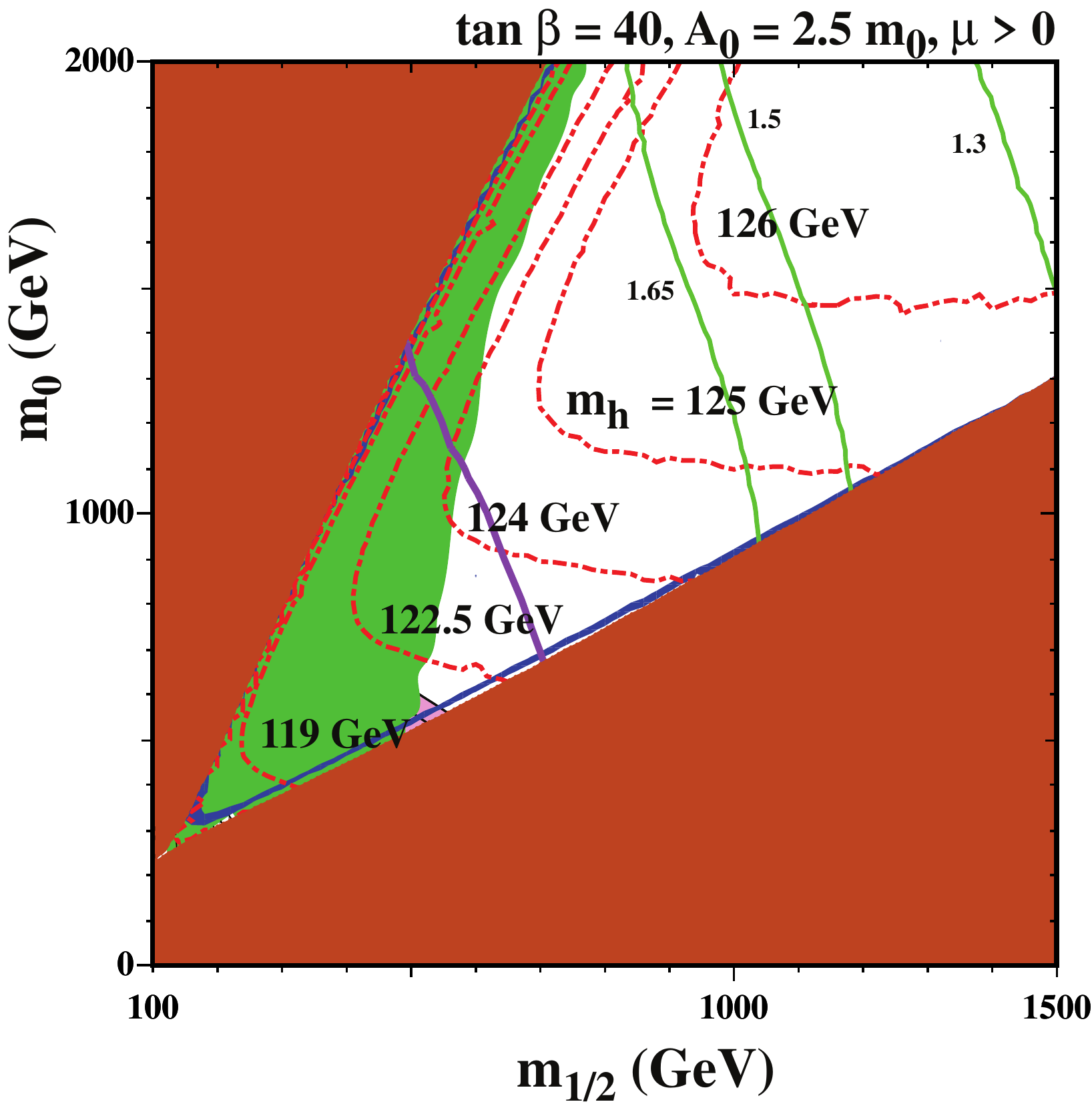}
\hfill
\end{minipage}\caption{
{\it
The CMSSM $(m_{1/2}, m_0)$ planes for $\mu > 0$, with
$\tan \beta = 10$ (left) and $40$ (right), $A_0 = 0$ (upper) and $A_0 = 2.5 \, m_0$ (lower), as 
calculated for $m_t = 173.2$~GeV using the latest version of the {\tt 
SSARD} code~\protect\cite{SSARD}. The interpretations of the shadings and contour colours
are described in the text.
}} 
\label{fig:CMSSM} 
\end{figure}

Fig.~\ref{fig:CMSSM} displays the interplays of these constraints in the $(m_{1/2}, m_0)$ planes of the
CMSSM for four choices of the input parameters $A_0$ and $\tan \beta$, assuming in all cases 
that $\mu > 0$~\footnote{Historically, this sign of $\mu$ has been favoured by the supersymmetric interpretation
of $g_\mu - 2$ (we display $g_\mu - 2$-compatible regions in Fig.~\ref{fig:CMSSM} but not in subsequent figures)
and to facilitate compatibility with $b \to s \gamma$. However, since the LHC
constraints now disfavour regions of the CMSSM parameter space that `explain' the $g_\mu - 2$
discrepancy (as seen in Fig.~\ref{fig:CMSSM}), this assumption should be taken with a grain of salt.}. The left panels of Fig.~\ref{fig:CMSSM}
are for $\tan \beta = 10$, and the right panels are for $\tan \beta = 40$. The upper panels are for $A_0 = 0$,
and the lower panels are for $A_0 = 2.5 \, m_0$. In each panel, the region at high $m_{1/2}$ and
low $m_0$ where the $\tilde{\tau}_1$ is the LSP is shaded brown, and the region at low $m_{1/2}$ and
high $m_0$ where there is no consistent solution to the vacuum conditions (\ref{eq:mu}, \ref{eq:muB}) is shaded mauve.
The regions excluded by $b \to s \gamma$ are shaded green, those favoured by $g_\mu - 2$ are shaded pink, and
those favoured by $\Omega_\chi h^2$ are shaded dark blue~\footnote{Because of the combination of recent experimental constraints, we are forced to consider substantially larger ranges in
the displayed supersymmetric parameters such as $m_0$ and $m_{1/2}$. Therefore,  here and
in many other figures, for reasons of visibility we shade
wider strips where $0.06 < \Omega_\chi h^2 < 0.2$.}. The LEP chargino exclusion
is shown as a near-vertical dashed black line at small $m_{1/2}$~\cite{LEPchargino}.

Turning to the LHC constraints, contours of $M_h$ are shown in all panels as red dash-dotted lines,
and the ATLAS exclusion from the search for events with missing transverse energy (MET) is shown
as a solid purple line~\footnote{As discussed in~\cite{mc8}, this limit is generally stronger than other direct
searches for supersymmetry at the LHC in the models studied, and is largely independent of $\tan \beta$
and $A_0$. Based on the analysis in~\cite{CELMOV}, in the upper left panel of Fig.~\ref{fig:CMSSM}
we have assumed that the constraint published in~\cite{ATLAS} is independent of $m_{1/2}$ when
extrapolated to lower $m_0$.}.  We also display as solid green lines three contours of 
BR$(B_s \to \mu^+ \mu^-)$/BR$(B_s \to \mu^+ \mu^-)_{SM} = 1.65, 1.5$ (the present 95\% CL upper limit from LHCb and combined experiments),
1.3 (the 68\% upper limit from LHCb).  For reference, where relevant we also show 
the contours for 1.0 (the Standard Model value) and 0.85 (corresponding to the central value
seen at LHCb).
The same shadings and line styles are used in all subsequent figures. 

Concerning the impacts of the LHC constraints, we see in the upper panels that
$M_h$ is in general too small if $A_0 = 0$. We also see that the MET constraint
allows only the upper end of the coannihilation strip close to the ${\tilde \tau_1}$
LSP boundary~\footnote{We recall that the focus-point strip is excluded by the
XENON100 upper limit on spin-independent dark matter scattering.}, and is
incompatible with a supersymmetric resolution of the $g_\mu - 2$ discrepancy. The
$B_s \to \mu^+ \mu^-$ constraint has no impact for $\tan \beta = 10$, but for
$\tan \beta = 40, A_0 = 0$ its restriction on the coannihilation strip is stronger than
that of the MET constraint.

In the lower panels, we see that $M_h$ is generally larger when $A_0 = 2.5 \, m_0$.
However, for $\tan \beta = 10$, 
$M_h$ does not grow above $\sim 121$ GeV, whereas for $\tan \beta = 40, A_0 = 2.5 \, m_0$ there
is compatibility for $m_{1/2} \gappeq 1$~TeV along the coannihilation strip. 
This region
is also compatible with the LHC MET constraint, but not with the supersymmetric
interpretation of $g_\mu - 2$.
Note that for the larger values of $A_0$, we see another brown shaded region in the upper
left where the stop becomes the LSP (or tachyonic). To the right of this boundary
there is a stop coannihilation strip which occurs at relatively low $M_h$ when $\tan \beta = 10$
and is in the region excluded by $b \to s \gamma$ when $\tan \beta = 40$. 
Once again, the $B_s \to \mu^+ \mu^-$ constraint has no impact for $\tan \beta = 10$,
and requires large $m_{1/2}$ for $\tan \beta = 40$. The disconnect between 
the LHC results and the potential solution to the $g_\mu - 2$ discrepancy 
is severe enough that we will not show the $g_\mu - 2$ compatible regions in future plots.

A more complete survey of CMSSM models with other values of $A_0$ is given in~\cite{eo6},
but this brief summary serves to illustrate how the LHC constraints prefer larger values of
$\tan \beta$, $A_0$ and $m_{1/2}$.

\subsection{Impacts on mSUGRA models}

\begin{figure}[htb!]
\begin{minipage}{8in}
\includegraphics[height=3.3in]{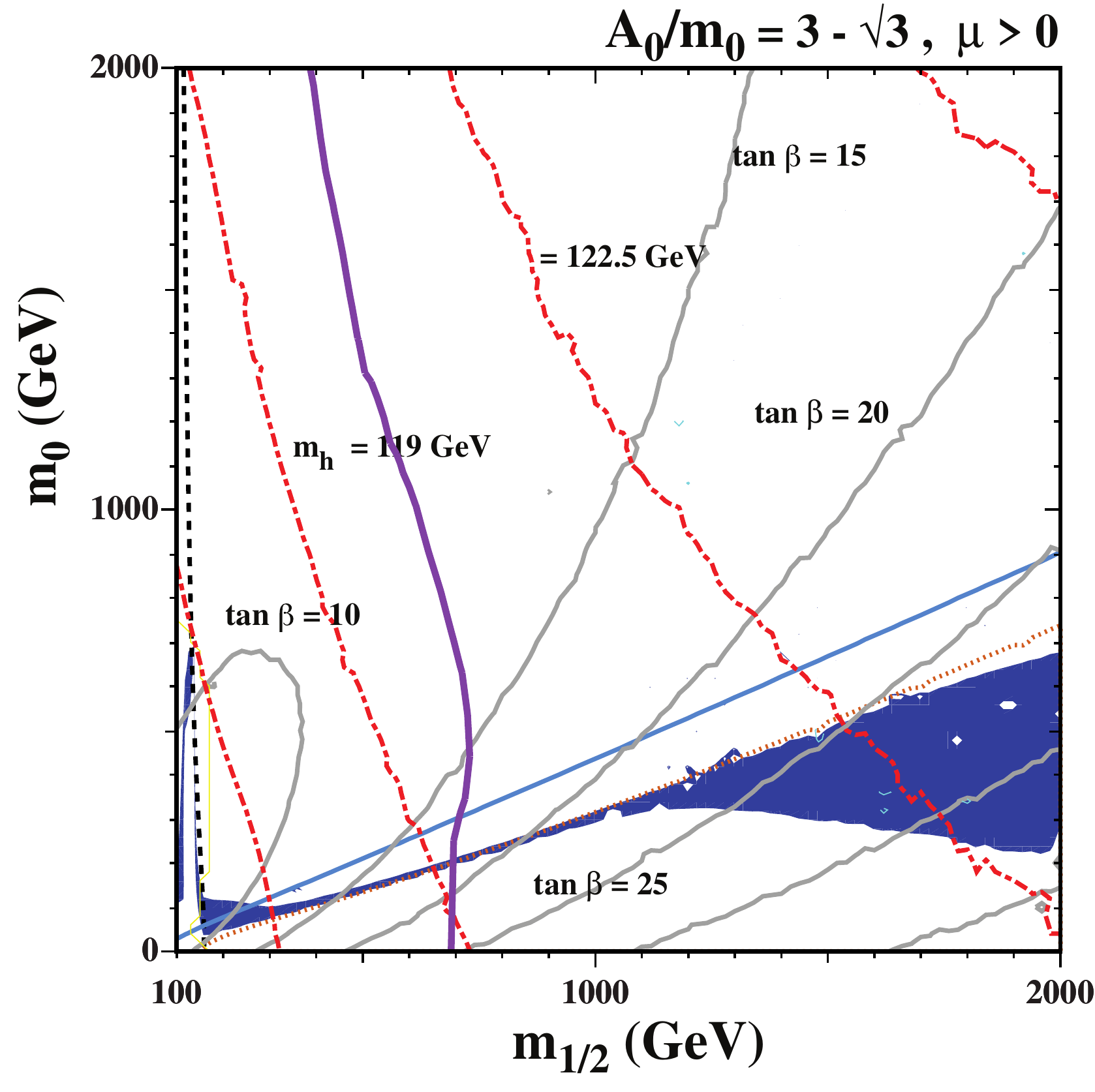}
\hspace*{-0.17in}
\includegraphics[height=3.3in]{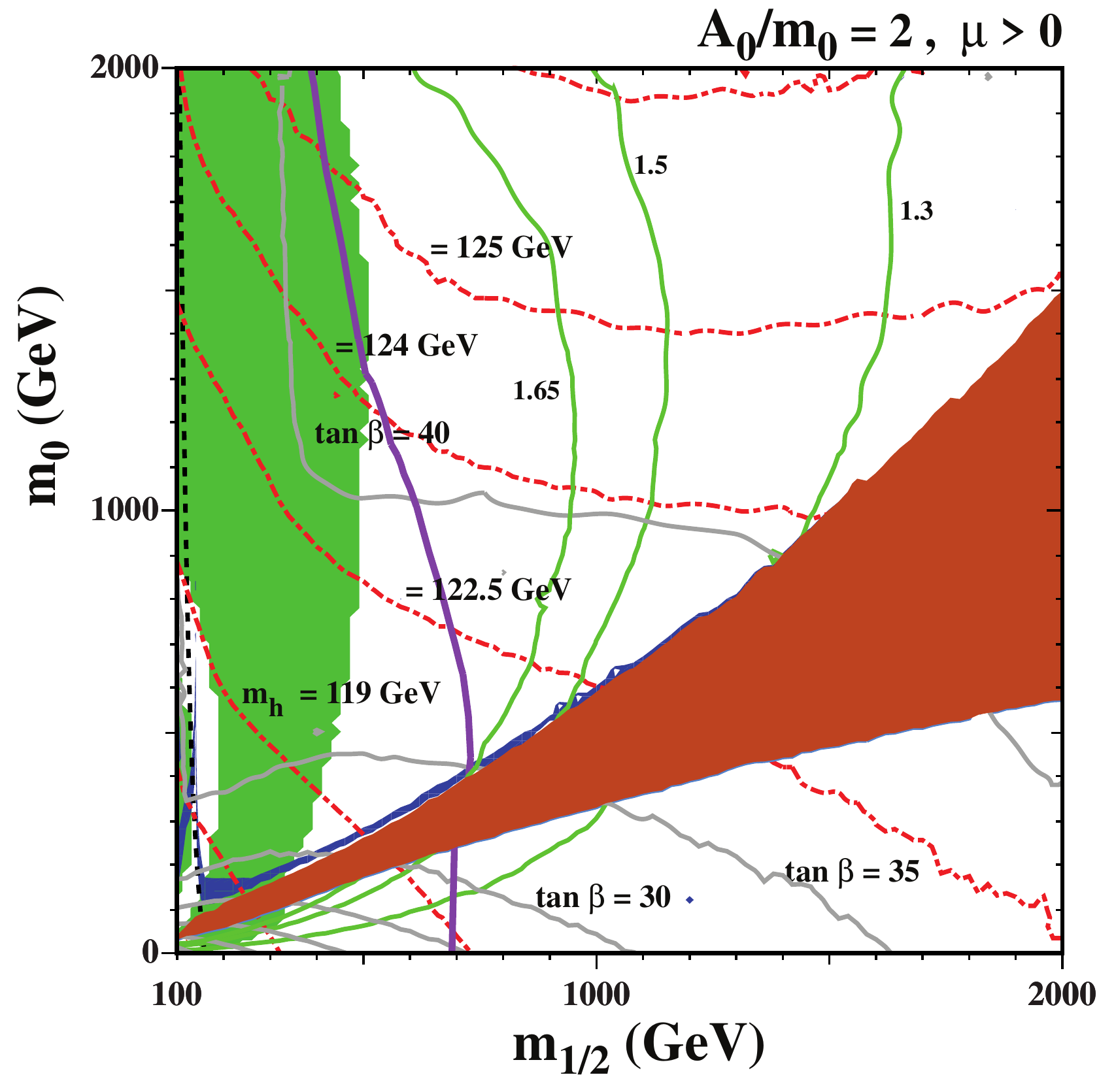}
\hfill
\end{minipage}
\caption{
{\it
The mSUGRA $(m_{1/2}, m_0)$ planes for $\mu > 0$
and $A_0 = (3-\sqrt{3})m_0$ (left), $A_0 = 2 m_0$ (right).
The interpretations of the shading and contour colours are described in the text.
}} 
\label{fig:msugra} 
\end{figure}

In Fig.~\ref{fig:msugra} we show a pair of mSUGRA $(m_{1/2}, m_0)$ planes for $\mu > 0$.
We recall that, there is just one free parameter in addition to these, which may be taken as
$A_0$: the electroweak
vacuum conditions (\ref{eq:mu}, \ref{eq:muB}) may be used to determine
$\tan \beta$ at each point on the plane~\cite{vcmssm}.
In the left panel, we adopt the value $A_0 = (3 - \sqrt{3}) \, m_0$ found in the original Polonyi model.
Contours of fixed $\tan \beta$ are shown by the grey solid curves in increments of 5, as labelled. 
Also shown, as a diagonal and solid light blue line, is the contour where $m_{3/2} = \min(m_\chi,m_{\tilde \tau_1})$,
below which the gravitino is the LSP.
Another diagonal line (brown dotted) shows the contour where the lightest neutralino mass 
$m_\chi$ is equal to the mass of the lighter stau,  $m_{\tilde \tau_1}$.
The latter appears below the gravitino LSP line, so the ${\tilde \tau_1}$
is never the LSP in this particular model. As a consequence, only the dark blue shaded region
at low $m_{1/2}$ above the light blue line corresponds to neutralino dark matter with the WMAP
density. The dark blue shaded region below the light blue line corresponds to the 
gravitino LSP having the WMAP density it would have if there is no non-thermal 
contribution to the gravitino density. Here, the gravitino density is determined
from the (co)annihilations of the NLSP - either the neutralino or (if below the dotted line) the lighter stau -
followed by decay of the NLSP into the gravitino LSP, resulting in 
$\Omega_{3/2} h^2 = (m_{3/2}/m_{\chi,{\tilde \tau_1}}) \Omega_{\chi,{\tilde \tau_1}} h^2$.
We recall that there are important cosmological and astrophysical constraints on
long-lived NLSP decays, which we do not discuss here.

Concerning the LHC constraints, we note that $M_h$ is always far below the LHC
measurement, which is a much stronger constraint than that from the MET search.
Both constraints are incompatible with the supersymmetric
interpretation of $g_\mu - 2$. Since $\tan \beta \le 35$ in the 
portion of the mSUGRA $(m_{1/2}, m_0)$ plane for $A_0 = (3 - \sqrt{3}) \, m_0$
displayed in the left panel of Fig.~\ref{fig:msugra},
the $B_s \to \mu^+ \mu^-$ constraint has no impact.

The right panel of Fig.~\ref{fig:msugra} shows a similar analysis for the case $A_0 = 2 \, m_0$.
Comparing with the left panel, we see that the values of $\tan \beta$ determined by the
electroweak vacuum conditions are generally larger. For this reason, there is a region at
low $m_{1/2}$ that is excluded by $b \to s \gamma$, and the $B_s \to \mu^+ \mu^-$ constraint
rules out a significant portion of the $(m_{1/2}, m_0)$ plane. We also see that the LHC
$M_h$ constraint is respected for $m_0 \gappeq 1000$~GeV.
In this case, there is a region where the stau is lighter than the lightest neutralino
and both are lighter than the gravitino: this ${\tilde \tau_1}$ LSP region is shaded dark brown.
In the lower right wedge at large $m_{1/2}$ and small $m_0$, below the brown excluded region, 
the gravitino is once again the LSP and the ${\tilde \tau_1}$ is the NLSP.
In this case, above the $\tilde{\tau}_1$ LSP region, we do find a WMAP co-annihilation strip, which extends to 
$m_{1/2} \sim 1100$ GeV. However, this strip has $M_h \lappeq 123$ GeV
and fails to respect the $B_s \to \mu^+ \mu^-$ constraint.

This summary serves to illustrate a broader incompatibility between the LHC constraints
and mSUGRA, traceable to the extra constraints in this model compared to the CMSSM.
We note that a more complete statistical analysis which included early LHC results already showed
 incompatibilities in the mSUGRA constructions \cite{mc4}.

\section{The NUHM}
 
\subsection{Results for NUHM1 Models}

\begin{figure}[htb!]
\begin{minipage}{8in}
\includegraphics[height=3.3in]{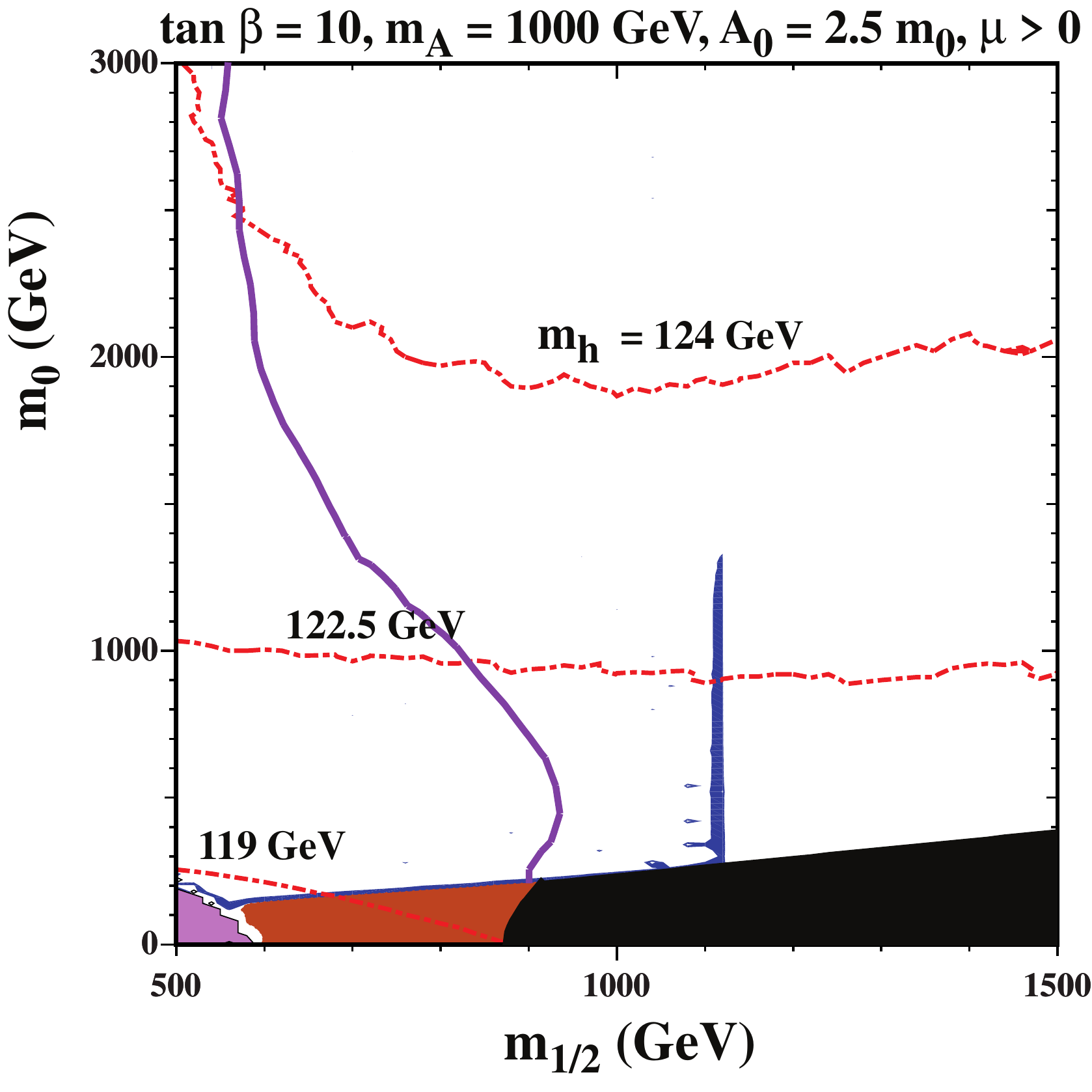}
\hspace*{-0.17in}
\includegraphics[height=3.3in]{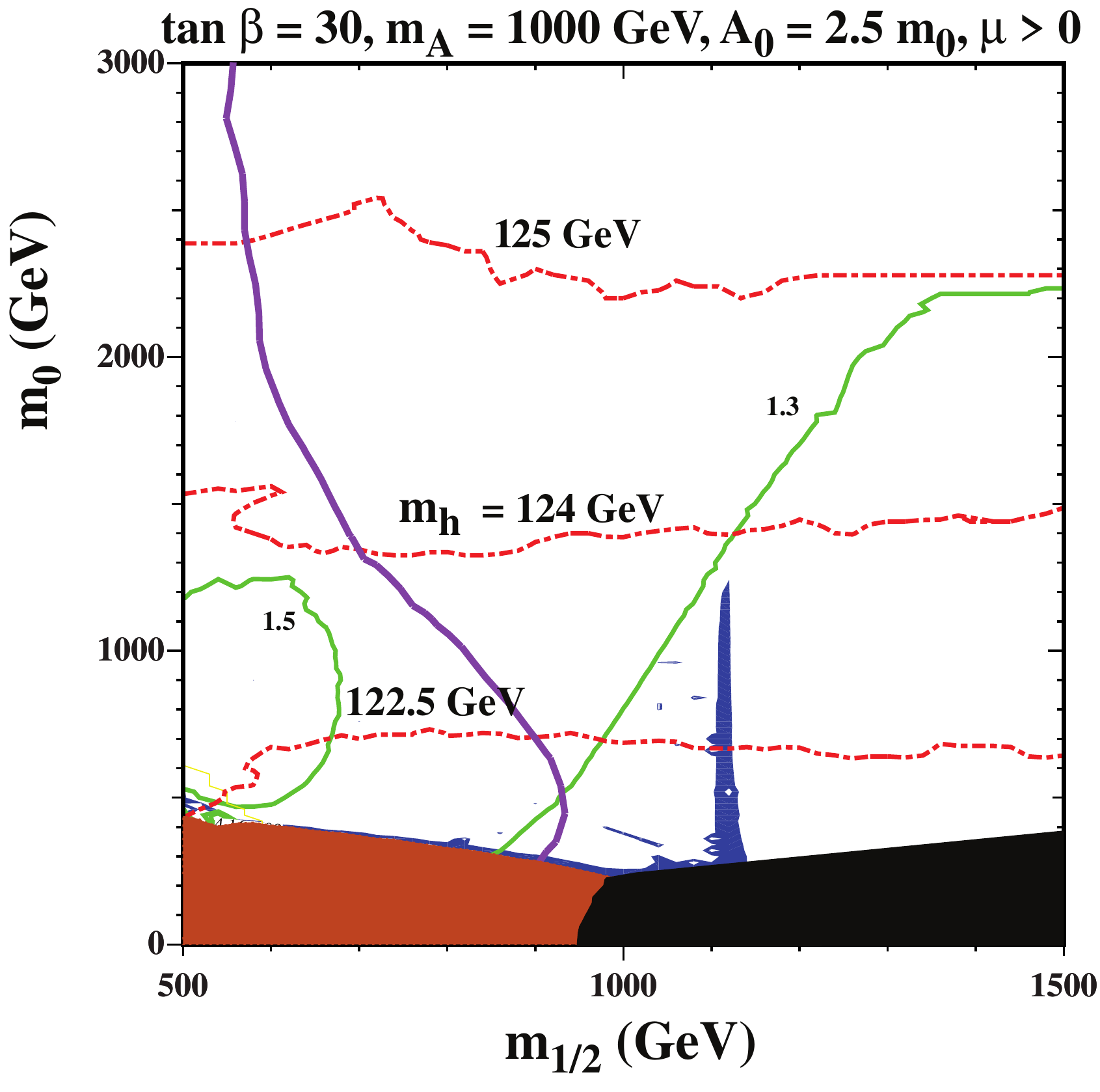}
\hfill
\end{minipage}
\caption{
{\it
The NUHM1 $(m_{1/2}, m_0)$ planes for $\mu > 0$, $m_A = 1000$~GeV, 
$A_0 = 2.5 \, m_0$ and $\tan \beta$ = 10 (left) and = 30 (right).
The interpretations of the shading and contour colours are described in the text.
}} 
\label{fig:nuhm1ma} 
\end{figure}

\begin{figure}[htb!]
\begin{minipage}{8in}
\includegraphics[height=3.3in]{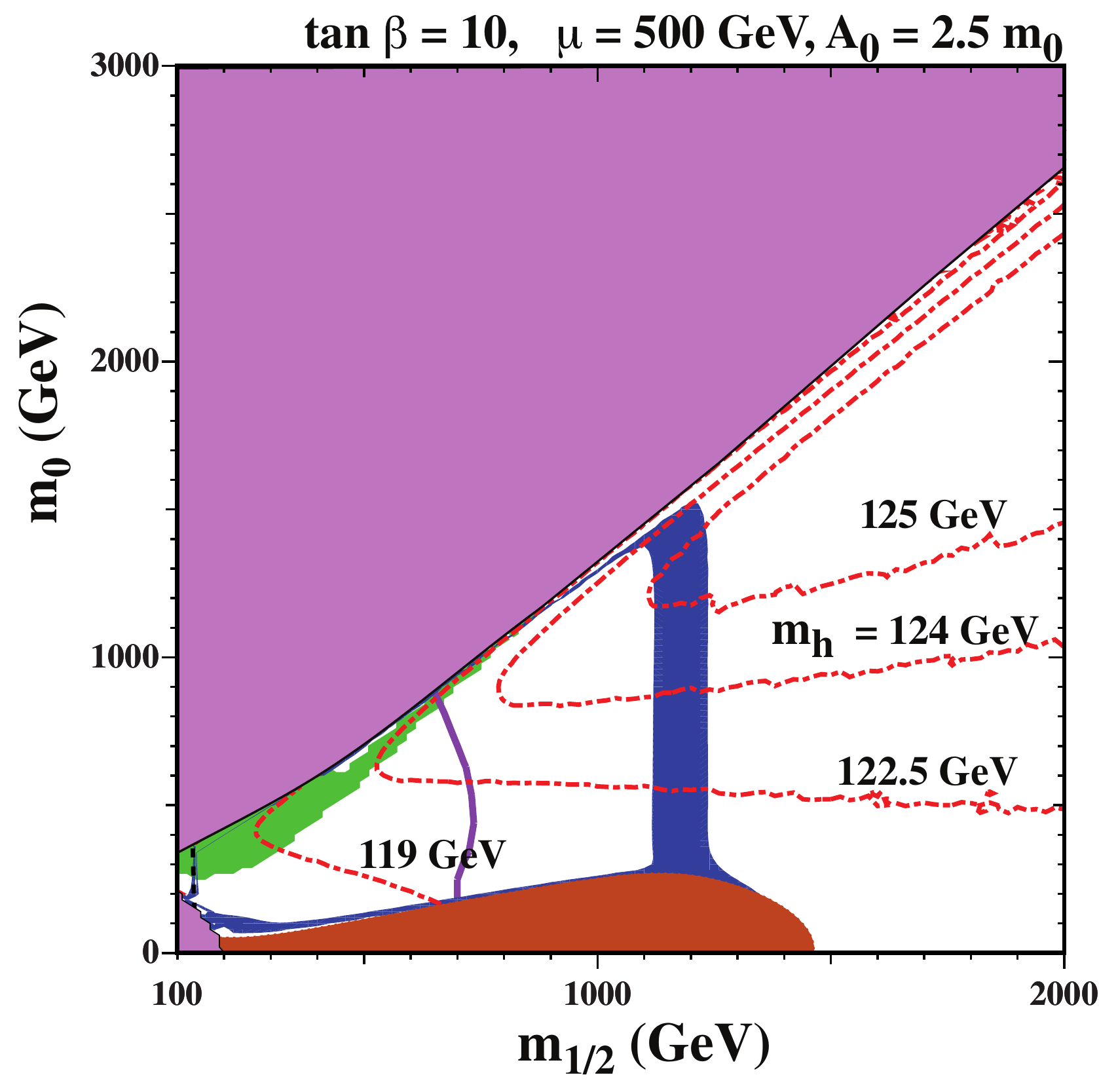}
\hspace*{-0.17in}
\includegraphics[height=3.3in]{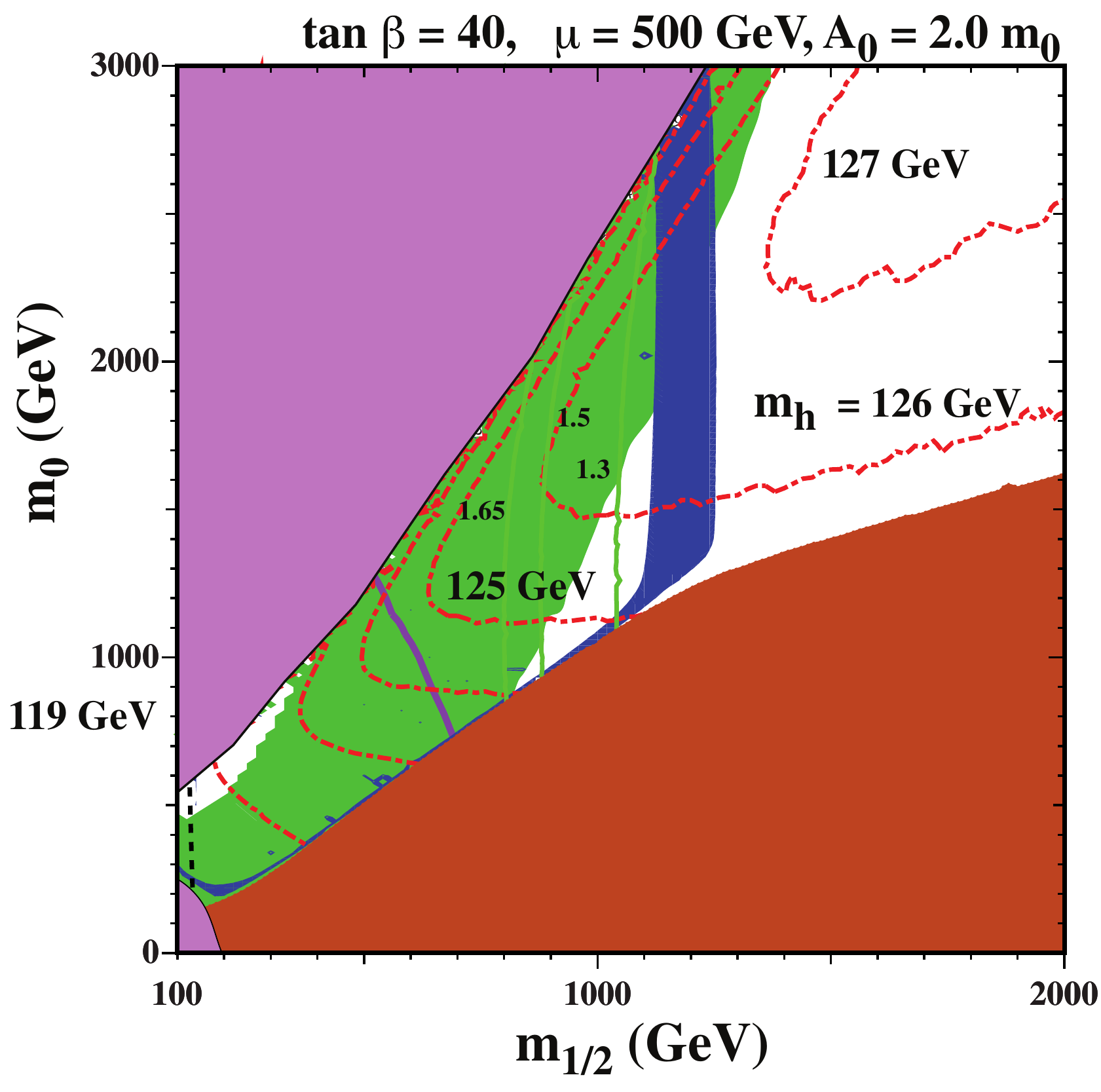}
\hfill
\end{minipage}
\caption{
{\it
The NUHM1 $(m_{1/2}, m_0)$ planes for $\mu > 0$, $\mu = 500$ GeV, 
 $\tan \beta$ = 10 and  $A_0 = 2.5 \, m_0$ (left) and $\tan \beta$ = 40 and  $A_0 = 2.0 \, m_0$ (right).
 The interpretations of the shading and contour colours are described in the text.
}} 
\label{fig:nuhm1mu} 
\end{figure}

We recall that the extra parameter introduced in the NUHM1 allows either $m_A$ or $\mu$
to be treated as a free parameter when solving the electroweak vacuum conditions 
(\ref{eq:mu}, \ref{eq:muB}) for any set of fixed values of $m_{1/2}, m_0, A_0$ and $\tan \beta$.
Fig.~\ref{fig:nuhm1ma} displays some representative $(m_{1/2}, m_0)$ planes for
fixed $A_0, \tan \beta$ and $m_A$, and Fig.~\ref{fig:nuhm1mu} displays analogous
planes for fixed values of $\mu$.

When $m_A$ is held fixed as in Fig. \ref{fig:nuhm1ma}, one finds 
an s-channel rapid annihilation funnel at a value of $m_{1/2}$ slightly greater than
$m_A$ (we recall that the funnel appears when $2 m_\chi \simeq m_A$, and that $m_\chi \sim
0.43 m_{1/2}$). This is seen as a vertical structure at $m_{1/2} \ga 1100$ GeV~\footnote{The width of the structure
is enhanced for visibility by colouring the area where $0.06 < \Omega_\chi h^2 < 0.2$.} that is
comfortably compatible with the LHC MET searches.
For $\tan \beta  = 10$, the funnel does not extend past $M_h \simeq 123$ GeV, whereas 
it extends almost up to $M_h = 124$ GeV for $\tan \beta = 30$. In both cases, 
the Higgs mass is probably too small to be compatible with the LHC result.
Again, on the other hand, since the funnels shown here were chosen to occur at relatively large $m_{1/2}$, 
there is no conflict with the LHC MET searches.  For $\tan \beta = 30$,
the branching ratio of $B_s \rightarrow \mu^+ \mu^-$ is somewhat higher
than the Standard Model value, but it is still within the current 95\% CL upper limit.
At $\tan \beta = 40$ (not displayed), there is increased tension with $B_s \rightarrow \mu^+ \mu^-$
as the funnel sits very close to the 95\% CL upper limit (a ratio of 1.5). In addition, 
at the higher value of $\tan \beta$, the end point of the funnel rises up  only to 123 GeV.

Large regions of acceptable parameter space are found in the $(m_{1/2}, m_0)$
plane when $\mu$ is held fixed, as seen in Fig. \ref{fig:nuhm1mu} where $
\mu = 500$ GeV. Here we have again fixed $\tan \beta = 10$ and $A_0 = 2.5 m_0$
in the left panel, but have taken $\tan \beta = 40$ and $A_0 = 2 m_0$ in the right panel.
For larger $A_0/m_0$, the $\tilde{\tau}_1$ LSP region (shaded brown) collides with the region
(shaded mauve) where there is no radiative electroweak symmetry breaking.
At low $m_{1/2}$, there is a $\tilde{\tau}_1$ coannihilation strip
which has low $M_h$ when $\tan \beta = 10$, and is for the most
part in conflict with $b \to s \gamma$ when $\tan \beta = 40$. 
At larger $m_{1/2}$, there is an increasing 
Higgsino component in the LSP and, as result, an increased annihilation cross section.
This results in the relatively thick vertical strips seen at $m_{1/2} \sim 1200$ GeV in both panels,
again comfortably consistent with the LHC MET searches,
where the transition towards a Higgsino-like LSP causes the relic density to fall into
the range favoured by WMAP.  At still larger $m_{1/2}$, the relic $\chi$ density is too small, but 
this region is not excluded if there is another dark matter candidate.

For the fixed parameter choices in Fig. \ref{fig:nuhm1mu}, this transition strip
lies beyond the current LHC MET search.  In this case, 
it is possible to obtain a Higgs mass in excess of 125 GeV even for $\tan \beta = 10$, 
and much of the transition strip lies in excess of $M_h = 126$ when $\tan \beta = 40$. 
As seen from the contours showing the branching ratio of $B_s \rightarrow \mu^+ \mu^-$,
this region is also compatible with the recent LHCb result.

\begin{figure}[htb!]
\begin{minipage}{8in}
\includegraphics[height=3.3in]{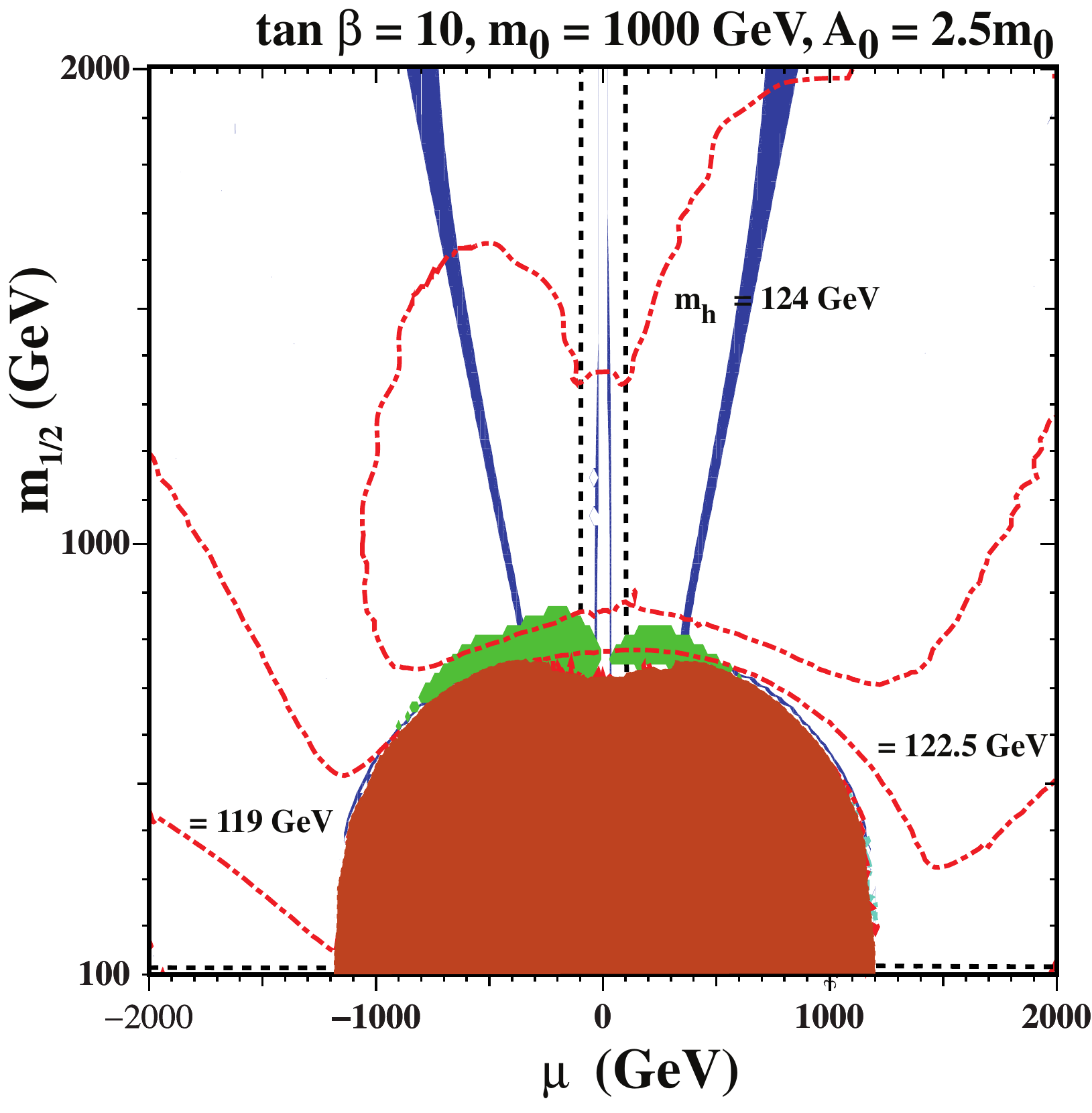}
\hspace*{-0.17in}
\includegraphics[height=3.3in]{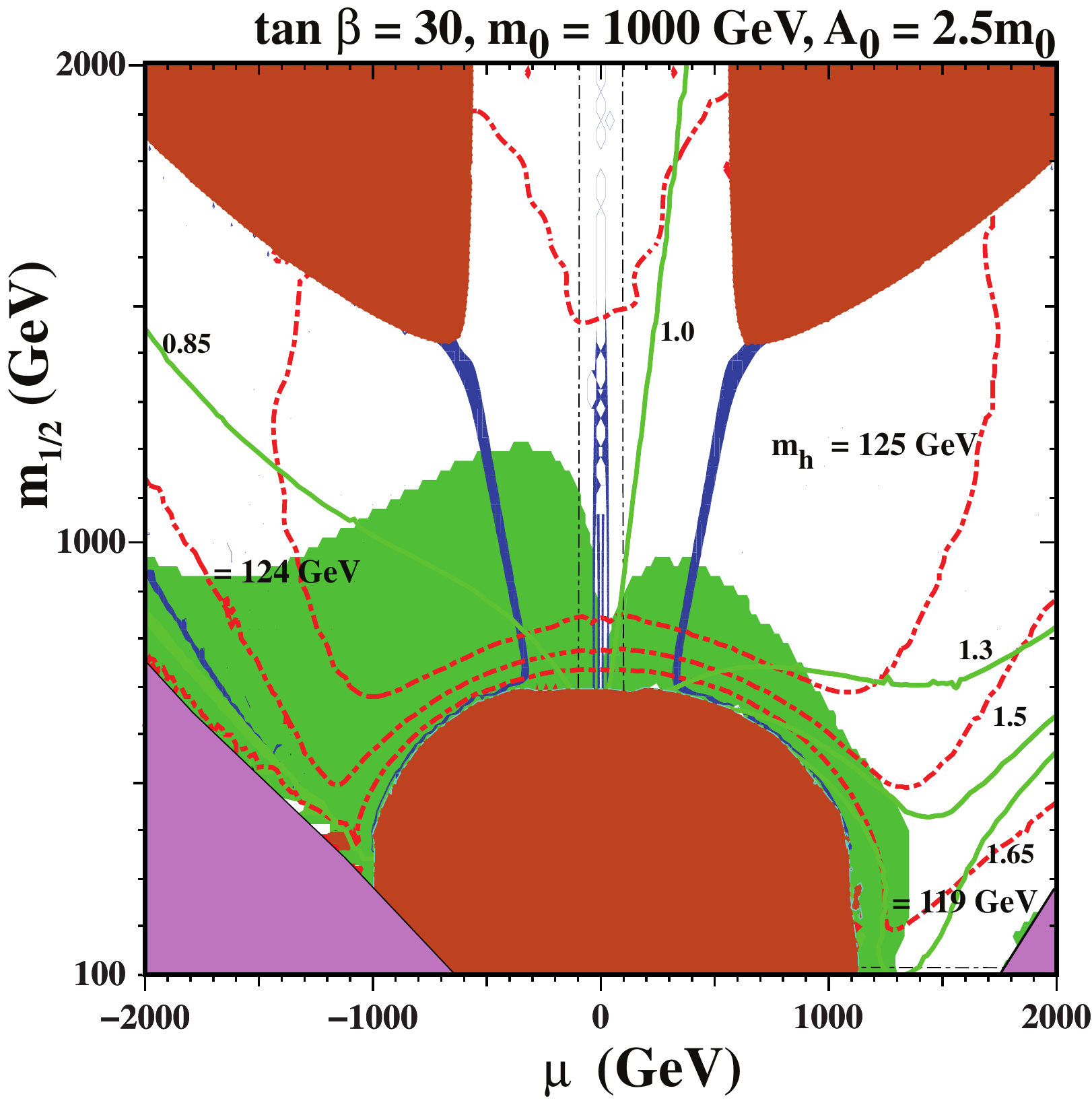}
\hfill
\end{minipage}
\caption{
{\it
The NUHM1 $(\mu, m_{1/2})$ planes for $m_0 = 1000$ GeV, $A_0 = 2.5 \, m_0$, and
 $\tan \beta$ = 10 (left) and $\tan \beta$ = 30 (right).
 The interpretations of the shading and contour colours are described in the text.
}} 
\label{fig:nuhm1mu2} 
\end{figure}

There are of course many possible slices that one can make through the NUHM1
parameter space, and we only show a few of the more interesting ones here.
In Fig.~\ref{fig:nuhm1mu2}, we show examples of $(\mu, m_{1/2})$ planes with
fixed $m_0 = 1000$ GeV and $A_0 = 2.5 \, m_0$, $\tan \beta = 10$ on the left and 
$\tan \beta = 30$ on the right. 
In both panels there is a brown shaded region at low $m_{1/2}$ and small $|\mu|$
corresponding to a stop LSP (or even a tachyonic stop). Just above the shaded
region, there is a thin strip where the relic density falls in the WMAP range due
to stop co-annihilations. However, if it is not excluded by $b \to s \gamma$, 
the value of $M_h$ is too low. Jutting out from this region, are two antenna-like
strips with $m_{1/2} \gappeq 800$~GeV and hence compatible with the LHC MET constraint,
which correspond again to the transition between bino and Higgsino-like
LSPs we met in the previous figure.  In the region between the two antennae, the relic density is too small.
For $\tan \beta$ = 10, $M_h$ is in excess of 124 GeV, and for $\tan \beta = 30$, 
it is in excess of 125 GeV.    For $\tan \beta = 30$, the brown shaded regions 
at large $m_{1/2}$ depict regions with a $\tilde{\tau}_1$ LSP. At still higher $\tan \beta$,
these would descend and cover much of the plane. 

It is interesting to note that in the right panel of Fig.~\ref{fig:nuhm1mu2}
the branching ratio of $B_s \to \mu^+ \mu^-$ falls below the Standard Model value, 
particularly when $\mu < 0$. Here we 
find a region with acceptable relic LSP density, a Higgs mass $M_h \simeq 125.5$ GeV,
acceptable $b \to s \gamma$, and $B_s \to \mu^+ \mu^-$ at 0.9 its Standard Model value,
close to the central value measured by LHCb.

\begin{figure}[htb!]
\begin{minipage}{8in}
\includegraphics[height=3.3in]{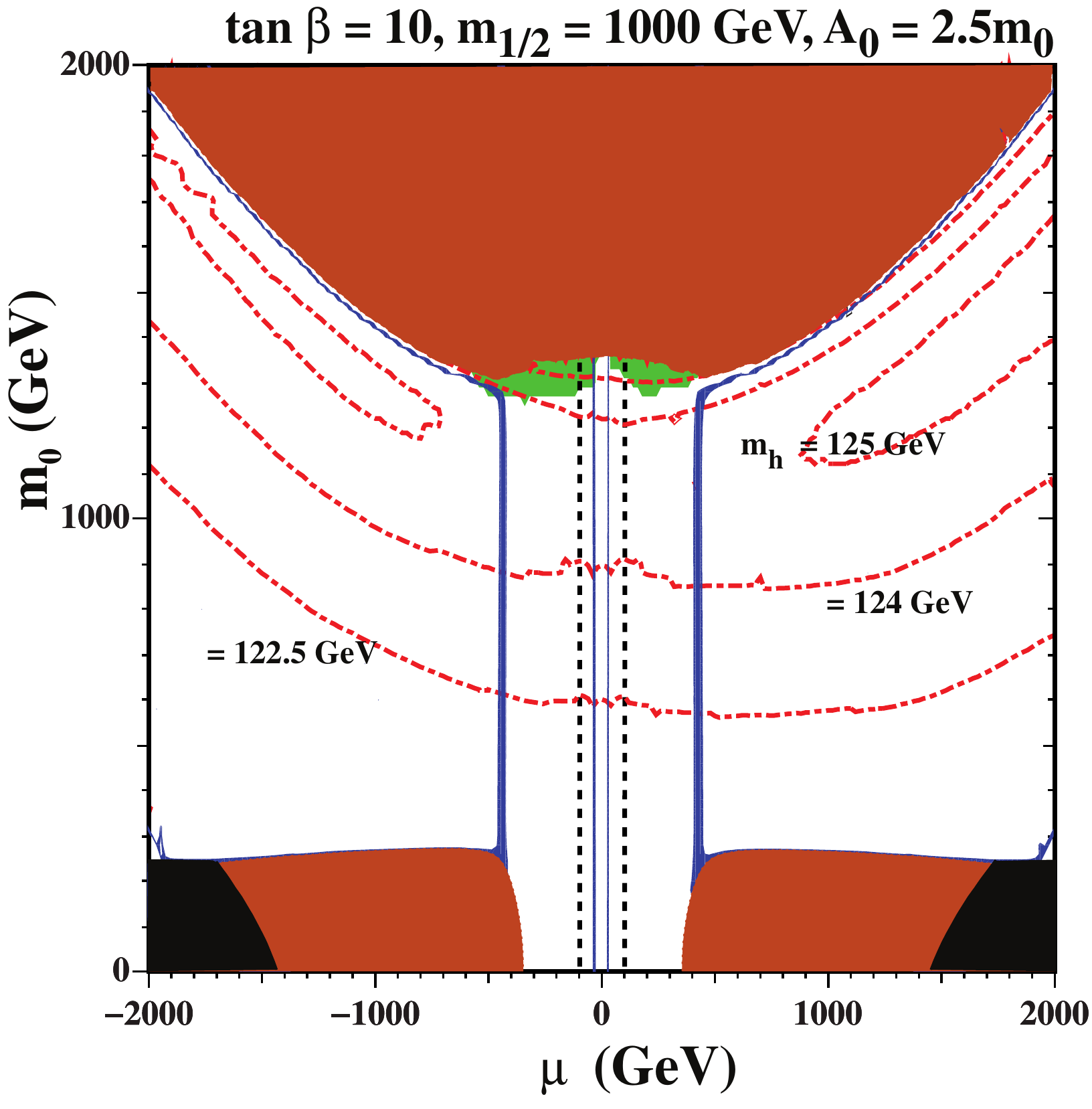}
\hspace*{-0.17in}
\includegraphics[height=3.3in]{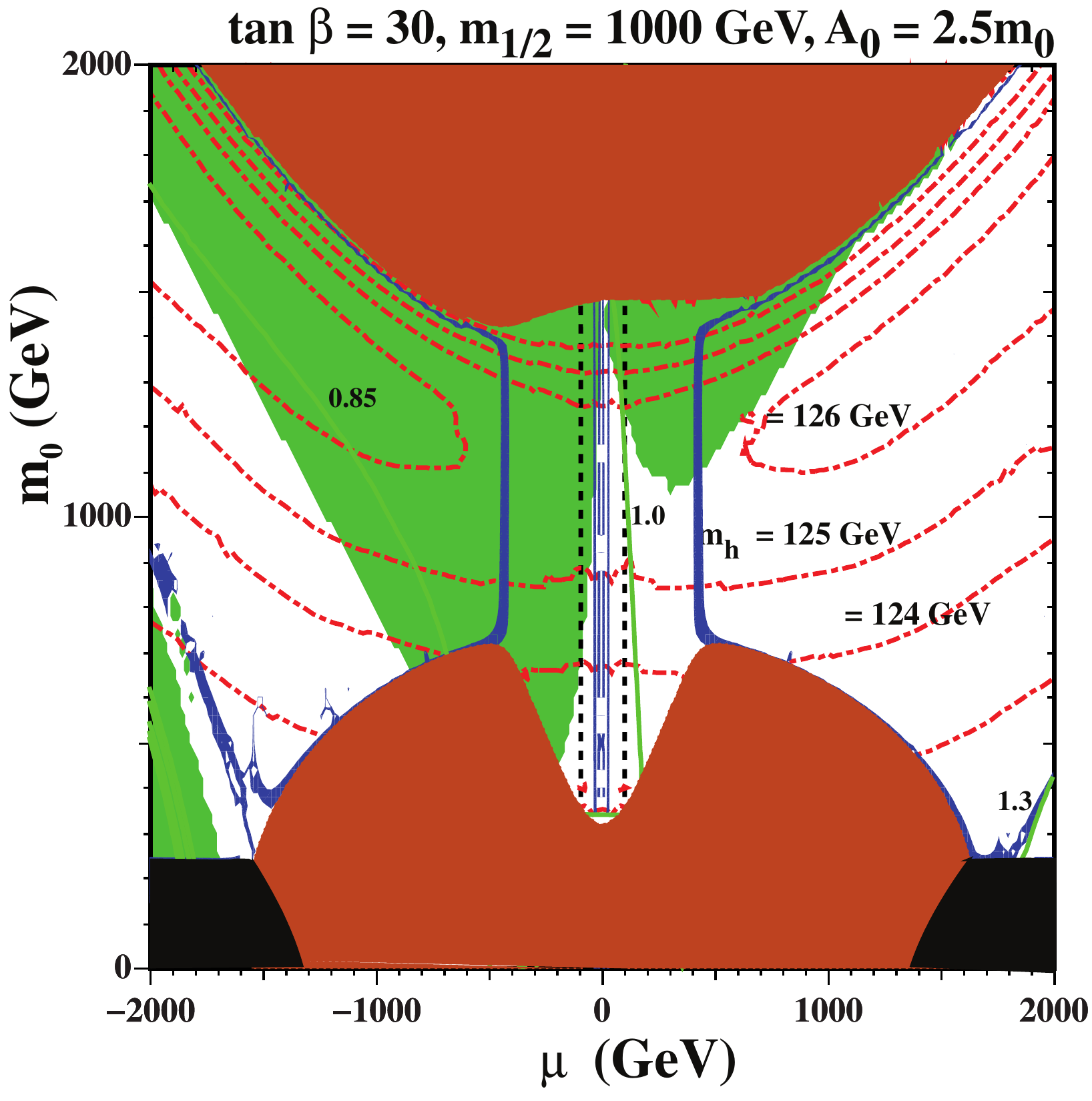}
\hfill
\end{minipage}
\caption{
{\it
The NUHM1 $(\mu, m_0)$ planes for $m_{1/2} = 1000$ GeV, $A_0 = 2.5 \, m_0$, and
 $\tan \beta$ = 10 (left) and $\tan \beta$ = 30  (right).
 The interpretations of the shading and contour colours are described in the text.
}} 
\label{fig:nuhm1mu3} 
\end{figure}

In Fig. \ref{fig:nuhm1mu3} we show one more example of NUHM1 planes, 
a pair of $(\mu, m_0)$ planes with fixed $m_{1/2} = 1000$~GeV
(and hence LHC MET-compatible) and $A_0 = 2.5 \, m_0$,
with $\tan \beta = 10$ on the left and $\tan \beta = 30$ on the right.
We again find regions with either a $\tilde{t}_1$ LSP (high $m_0$) or a $\tilde{\tau}_1$ LSP (low $m_0$)
with corresponding adjacent $\tilde{t}_1$ or $\tilde{\tau}_1$ coannihilation strips. These coannihilation strips
have either a low Higgs mass or are in conflict with $b \to s \gamma$. 
At larger values of $|\mu|$ we also see regions where
sleptons of the first two generations can become the LSP (coloured black). 
In both panels, there are  vertical transition strips with suitably 
large values of $M_h$. The constraint from $b \to s \gamma$ 
becomes particularly strong for $\mu < 0$ and $\tan \beta = 30$, while
the transition strip for $\mu > 0$ has a branching fraction for $B_s \to \mu^+ \mu^-$
close to the Standard Model prediction.

These examples illustrate that the NUHM1 has the flexibility to accommodate
the LHC constraints in a quite generic way, exploiting the extra flexibility in the
Higgs sector of this model.

\subsection{Results for NUHM2 Models}

There are still more planes that can be constructed 
in the NUHM2, with its extra free parameter. Here we show only one pair of $(\mu, m_A)$ planes
that are LHC MET-compatible,
with fixed $m_{1/2} = m_0 = 1000$ GeV and $A_0 = 2.5 m_0$ and, as in previous
planes, $\tan \beta = 10$ on the left and $\tan \beta = 30$ on the right.
In both these particular examples, we see cruciform regions of acceptable
relic density formed by the combination of a funnel, around fixed $m_A \simeq 900$ GeV,
and a transition to a Higgsino LSP when $|\mu| \la 500$ GeV.
The constraint from $b \to s \gamma$ disfavours more of the parameter space
with $\mu < 0$, particularly when $\tan \beta = 30$. For this value of $\tan \beta$,
the constraint from $B_s \to \mu^+ \mu^-$ excludes the surviving portions of the
relic density strips below $m_A \approx 1000$ GeV when $\mu > 0$. However,
at large $m_A$,  BR($B_s \to \mu^+ \mu^-$) is sufficiently small and $M_h \simeq 
125$~GeV. In the case of $\tan \beta = 10$, $M_h \gappeq 124$~GeV for most of
the cruciform region of acceptable relic density, and $B_s \to \mu^+ \mu^-$ has no
significant impact.

\begin{figure}[htb!]
\begin{minipage}{8in}
\includegraphics[height=3.3in]{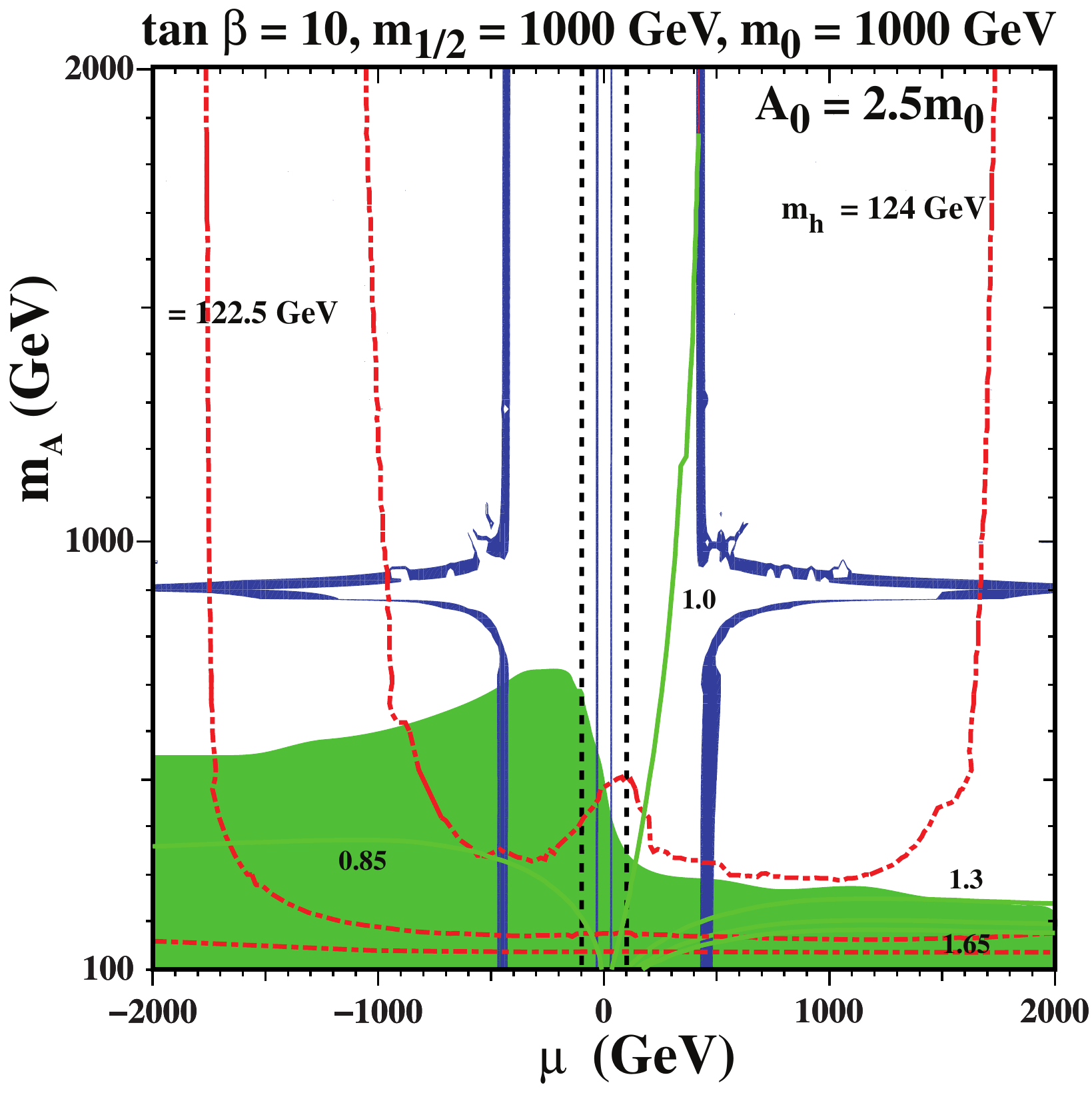}
\hspace*{-0.17in}
\includegraphics[height=3.3in]{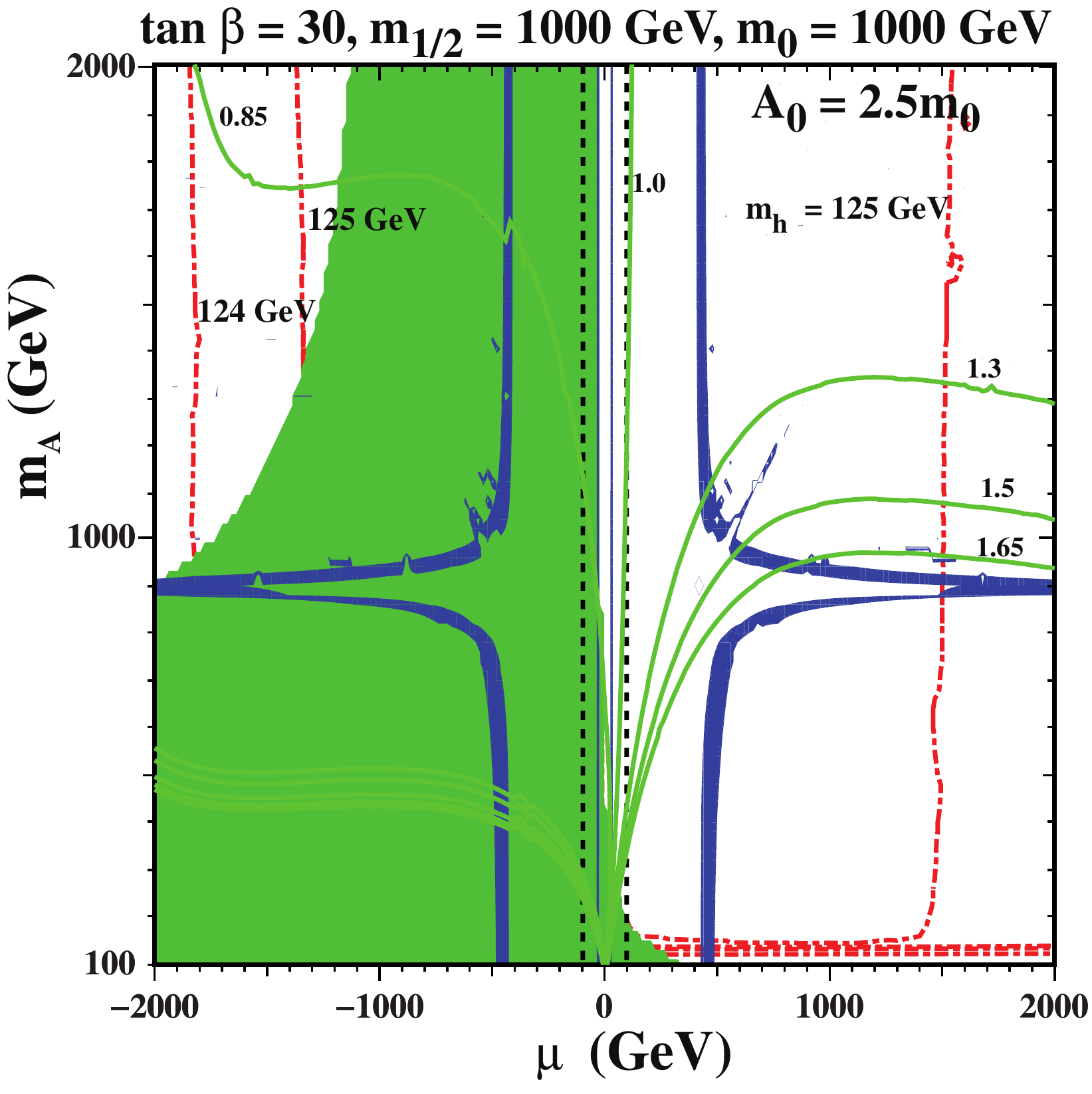}
\hfill
\end{minipage}
\caption{
{\it
The NUHM2 $(\mu, m_A)$ planes for $m_{1/2}  = 1000$ GeV, $m_0 = 1000$ GeV, 
$A_0 = 2.5 \, m_0$, and 
 $\tan \beta$ = 10   (left) and $\tan \beta$ = 30 (right).
 The interpretations of the shading and contour colours are described in the text.
}} 
\label{fig:nuhm2} 
\end{figure}

These two examples illustrate that the extra degree of freedom in the NUHM2
opens up even broader possibilities for reconciling supersymmetry with the
LHC constraints.

\section{Results for sub-GUT CMSSM models}

We now consider a different one-parameter extension, one in which the
soft supersymmetry-breaking parameters $m_{1/2}, m_0$ and $A_0$ are assumed
to be universal at some input scale $M_{in}$ {\it below} the conventional supersymmetric GUT scale,
as might arise in models where the supersymmetry-breaking dynamics acts below
the GUT scale. As was discussed in~\cite{subGUT} and summarized earlier, the sparticle spectrum in such
a scenario is in general more compressed than in the conventional CMSSM
with, e.g., smaller mass differences between squarks and sleptons and between
different types of inos.  Models with similarly compressed spectra and the prospects for their discovery have been studied in~\cite{compressed}.

\begin{figure}[htb!]
\begin{minipage}{8in}
\includegraphics[height=3.3in]{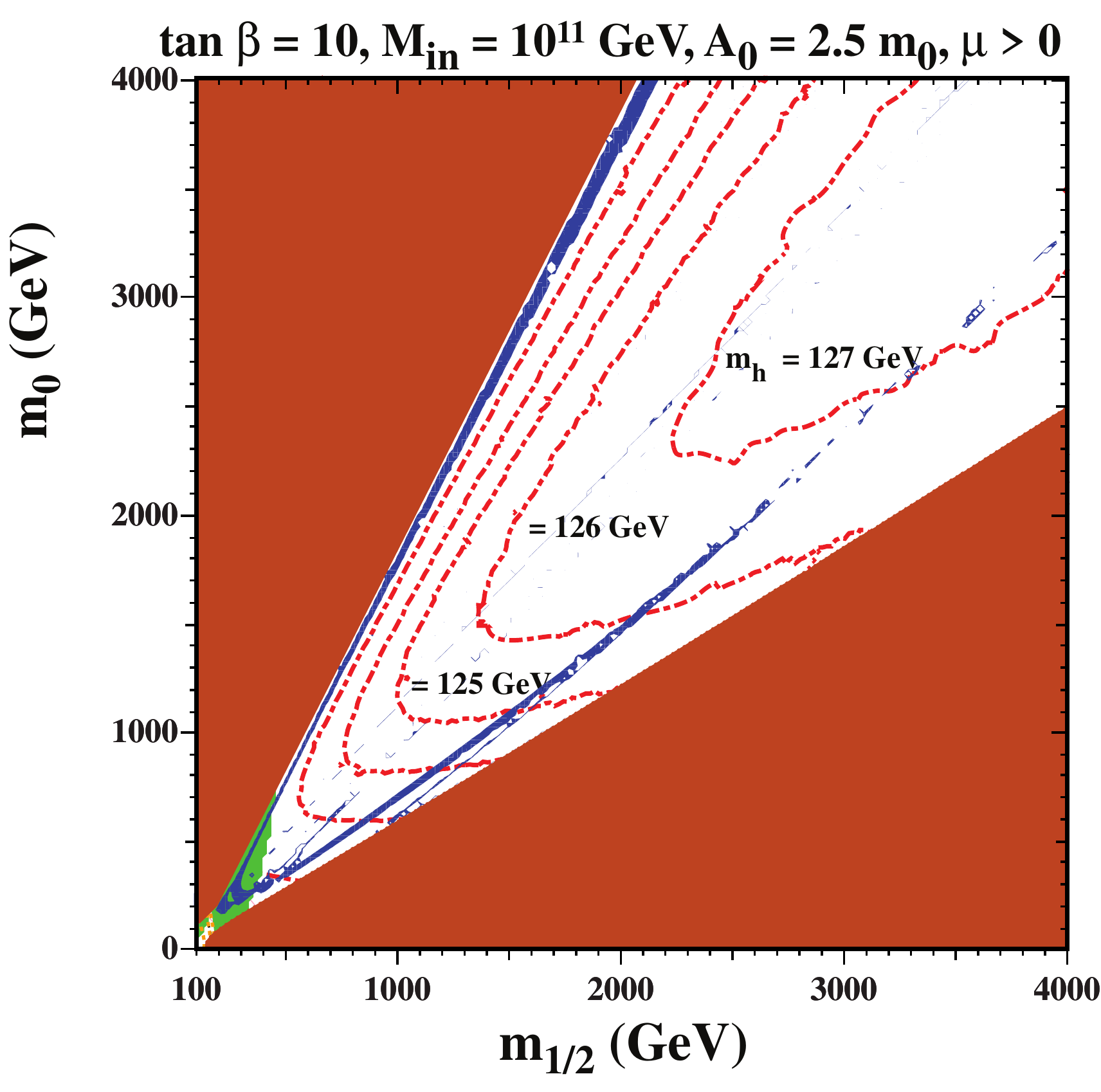}
\hspace*{-0.17in}
\includegraphics[height=3.3in]{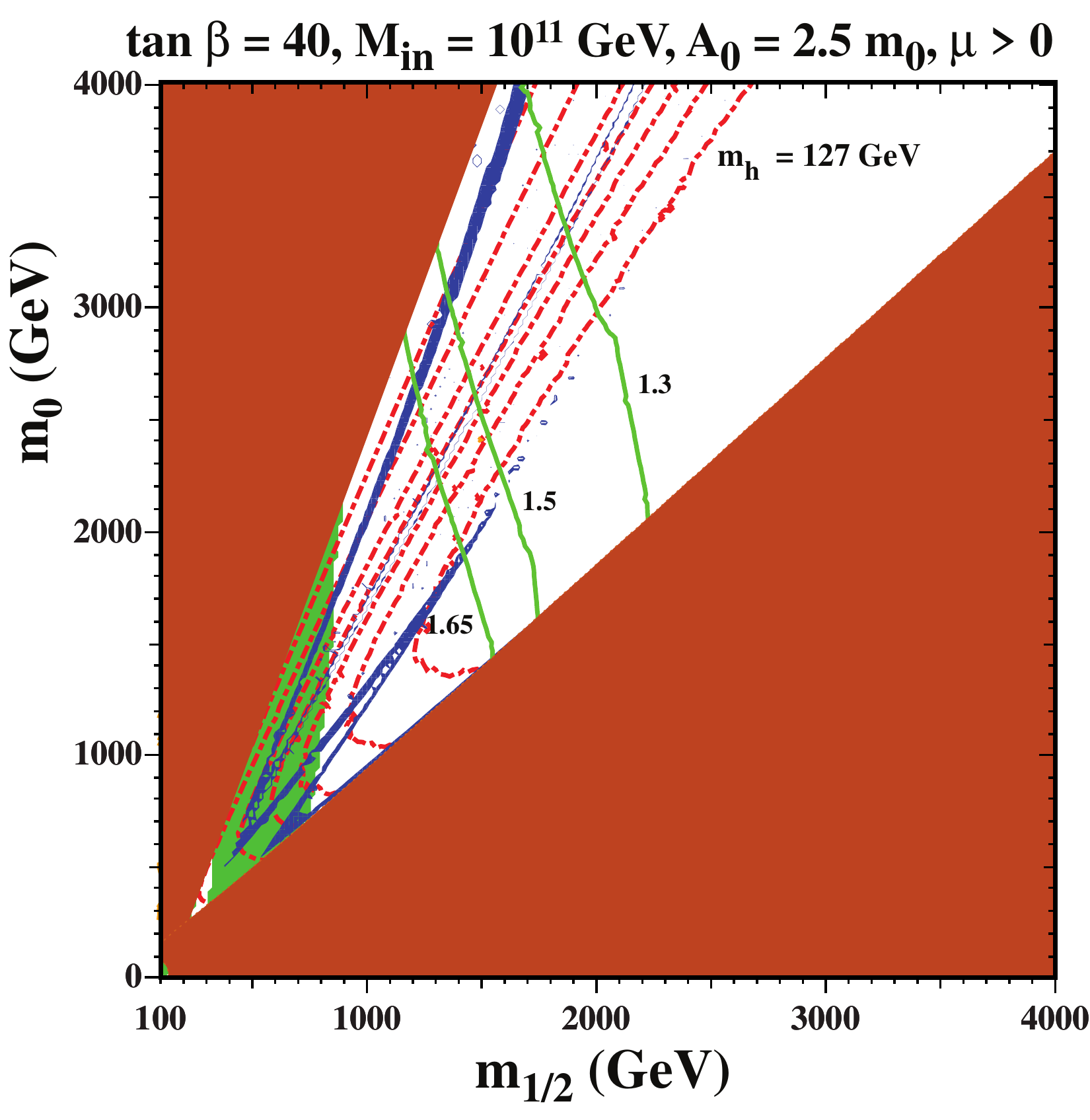}
\hfill
\end{minipage}
\begin{minipage}{8in}
\includegraphics[height=3.3in]{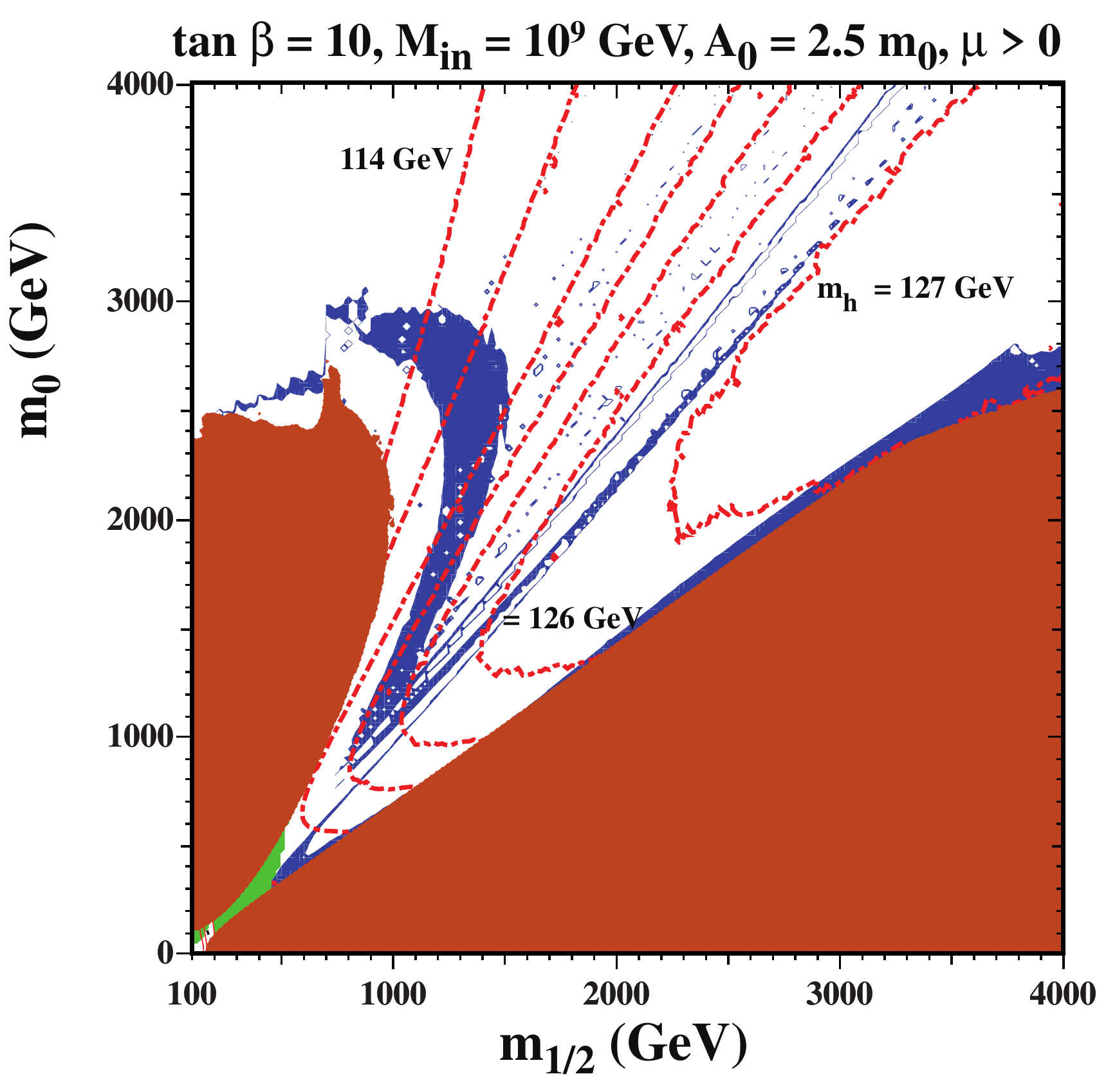}
\hspace*{-0.17in}
\includegraphics[height=3.3in]{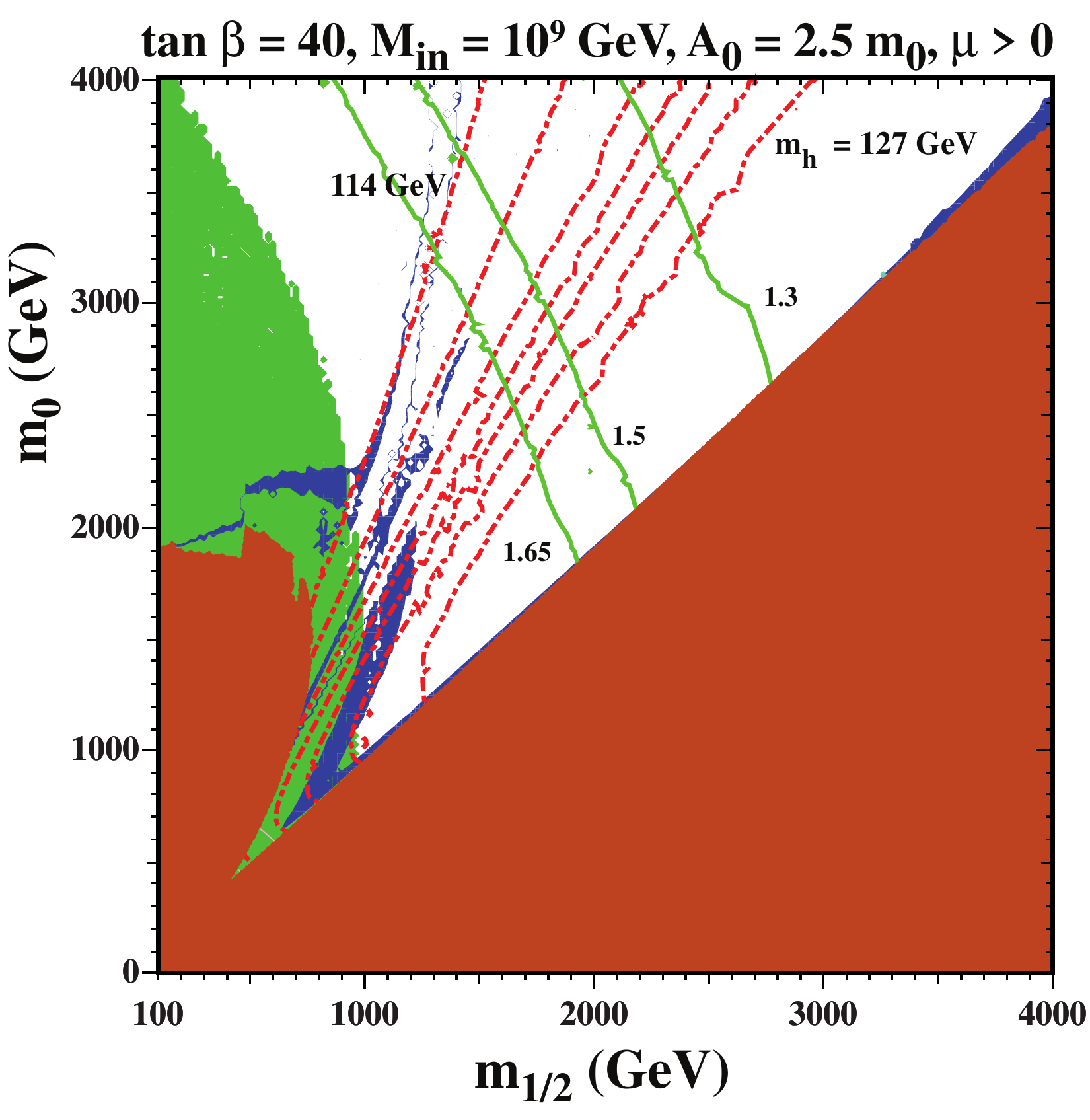}
\hfill
\end{minipage}\caption{
{\it
Sub-GUT CMSSM scenarios with $\tan \beta = 10$ (left) and $\tan \beta = 40$ (right)
in both cases with $A_0 = 2.5 \, m_0$, and for $M_{in} = 10^{11}$~GeV (upper)
and $10^9$~GeV (lower). Plotted contours of $M_h$ are 114, 119, 122.5, 124, 125,
126, and 127 GeV.
 The interpretations of the shading and contour colours are described in the text.
}} 
\label{fig:subGUT} 
\end{figure}

As presented in~\cite{subGUT}, this compression of the spectrum causes
the relic cold dark matter density to decrease at generic fixed values of
$(m_{1/2}, m_0)$ as $M_{in}$ decreases, and hence the strips of parameter space with the appropriate
dark matter density tend to move inwards from the boundary of the region allowed
by the neutral LSP and electroweak symmetry breaking constraints. These features
are visible in the upper panels of Fig.~\ref{fig:subGUT}, which are $(m_{1/2}, m_0)$
planes for $\tan \beta = 10$ and 40, both with $M_{in} = 10^{11}$~GeV,
$A_0 = 2.5 \, m_0$, and $\mu > 0$. In both cases there are wedges of the
$(m_{1/2}, m_0)$ plane at small $m_{1/2}$ and large $m_0$ (and hence large $A_0$) 
that are excluded
because the lighter stop, $\tilde{t}_1$, is the LSP in addition to the common wedge in the lower right of the planes
where there is a $\tilde{\tau}_1$ LSP. We see that in both panels there is a prominent funnel due to s-channel
annihilations of LSPs through the heavy Higgs scalar and pseudoscalar, where
the relic cold dark matter density is suitable. There are also other very thin strips running
above and roughly parallel to the prominent funnels, due to rapid s-channel {\em coannihilation} processes between the LSP and heavier neutralinos and charginos.
In both cases, we also see $\tilde{t}_1$ coannihilation strips
near the upper boundary of the allowed $\chi$ LSP region,
in the neighbourhood where focus-point strips appear in the CMSSM.
 
We now discuss how these sub-GUT models fare with the available LHC constraints.
The MET constraint (not shown) excludes only a small corner of the displayed triangle between
the $\tilde{\tau}_1$ and $\tilde{t}_1$ LSP boundaries.
As usual, $B_s \to \mu^+ \mu^-$ does not impact the allowed parameter space for
$\tan \beta = 10$, but does have significant impact for $\tan \beta = 40$, requiring
$m_{1/2} \gappeq 1500$ to 2000~GeV along the prominent rapid-annihilation
funnel, and $m_{1/2} \gappeq 1000$ to 1500~GeV along the stop coannihilation strip.
The most novel feature of these sub-GUT models is the relative ease with which they
respect the Higgs mass constraint. Most of the funnel for $\tan \beta = 10$ is
compatible with $M_h = 125$ to 126~GeV, once the uncertainties in the theoretical
calculation within the CMSSM are taken into account. This is also true along the funnel
when $\tan \beta = 40$, for the points compatible with $B_s \to \mu^+ \mu^-$. On the
other hand, in both cases the value of $M_h$ is too small along the stop coannihilation strip.

The lower panels of Fig.~\ref{fig:subGUT} display $(m_{1/2}, m_0)$
planes for the same values of $\tan \beta$, $\mu > 0$ and $A_0 = 2.5 \, m_0$,
but now with $M_{in} = 10^{9}$~GeV. In each case, rotating anticlockwise around the plane from the lower right,
we first see a stau coannihilation strip followed by a rapid-annihilation funnel
(which are both more prominent for $\tan \beta = 10$). In the $\tan \beta = 10$ case
(lower left) we also see `echo' funnel strips where $m_{1/2}$ is smaller for
the same value of $m_0$, which are due to s-channel coannihilations between the LSP
and more massive neutralinos and charginos. 
At lower $M_{in}$, there is less running and stops remain heavier than the
lightest neutralino at lower $m_0$, thus, we see a recession of the stop LSP region
at lower $M_{in}$. 
Next to this region, we see a prominent stop
coannihilation strip, but note that only for small $m_0$ does it have an acceptable
value of $M_h$. In the $\tan \beta = 40$ case, the s-channel funnels merge with the
stop coannihilation strip for $(m_{1/2}, m_0) \sim (1000, 2000)$~GeV
(adjacent to a region excluded by $b \to s \gamma$), again yielding
acceptable $M_h$ for lower $m_0$. The LHC MET constraint is not relevant in
either of the lower planes, and neither is $B_s \to \mu^+ \mu^-$ when $\tan \beta = 10$.
However, $B_s \to \mu^+ \mu^-$ does exclude almost all the dark matter regions for
$\tan \beta = 40$, with the exception of the stau coannihilation strip which reemerges
at very large $m_{1/2}$.

These examples illustrate how sub-GUT CMSSM models are better able to
accommodate the LHC constraints, principally because the greater importances
of additional (co)annihilation processes permit larger values of $m_{1/2}$ and $m_0$
than in the conventional GUT-scale CMSSM. 

\section{Results for sub-GUT mSUGRA models}

Encouraged by the relative success of sub-GUT CMSSM models in accommodating
the latest LHC constraints together with other constraints, we now consider whether
this compatibility can be maintained if we remove one of the sub-GUT CMSSM 
parameters by imposing the mSUGRA relation between the trilinear and bilinear
soft supersymmetry-breaking parameters: $A_0 = B_0 + m_0$, and also the mSUGRA
relation between the gravitino and scalar masses: $m_{3/2} = m_0$ before
renormalization.

\begin{figure}[htb!]
\begin{minipage}{8in}
\includegraphics[height=3.3in]{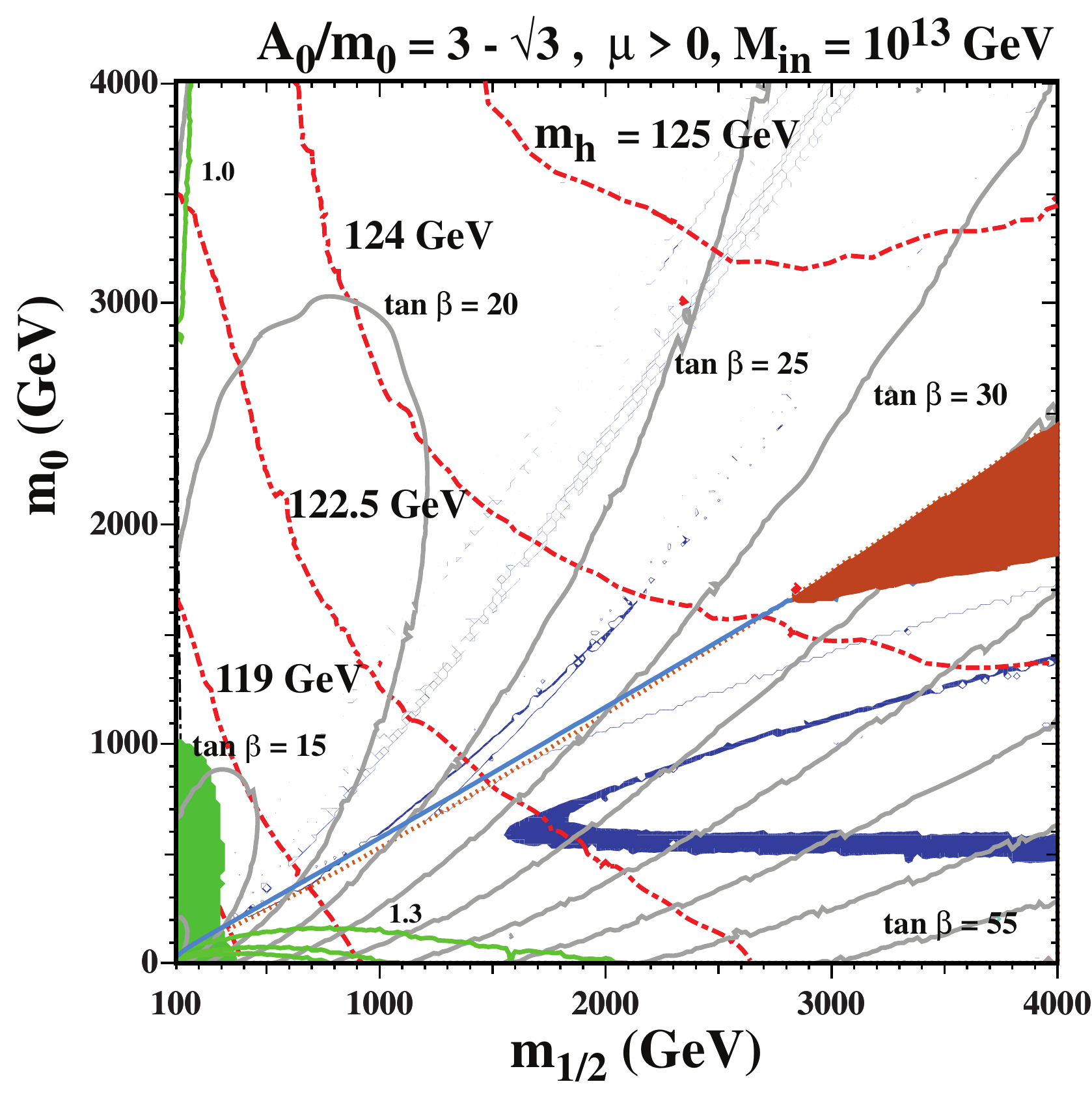}
\hspace*{-0.17in}
\includegraphics[height=3.3in]{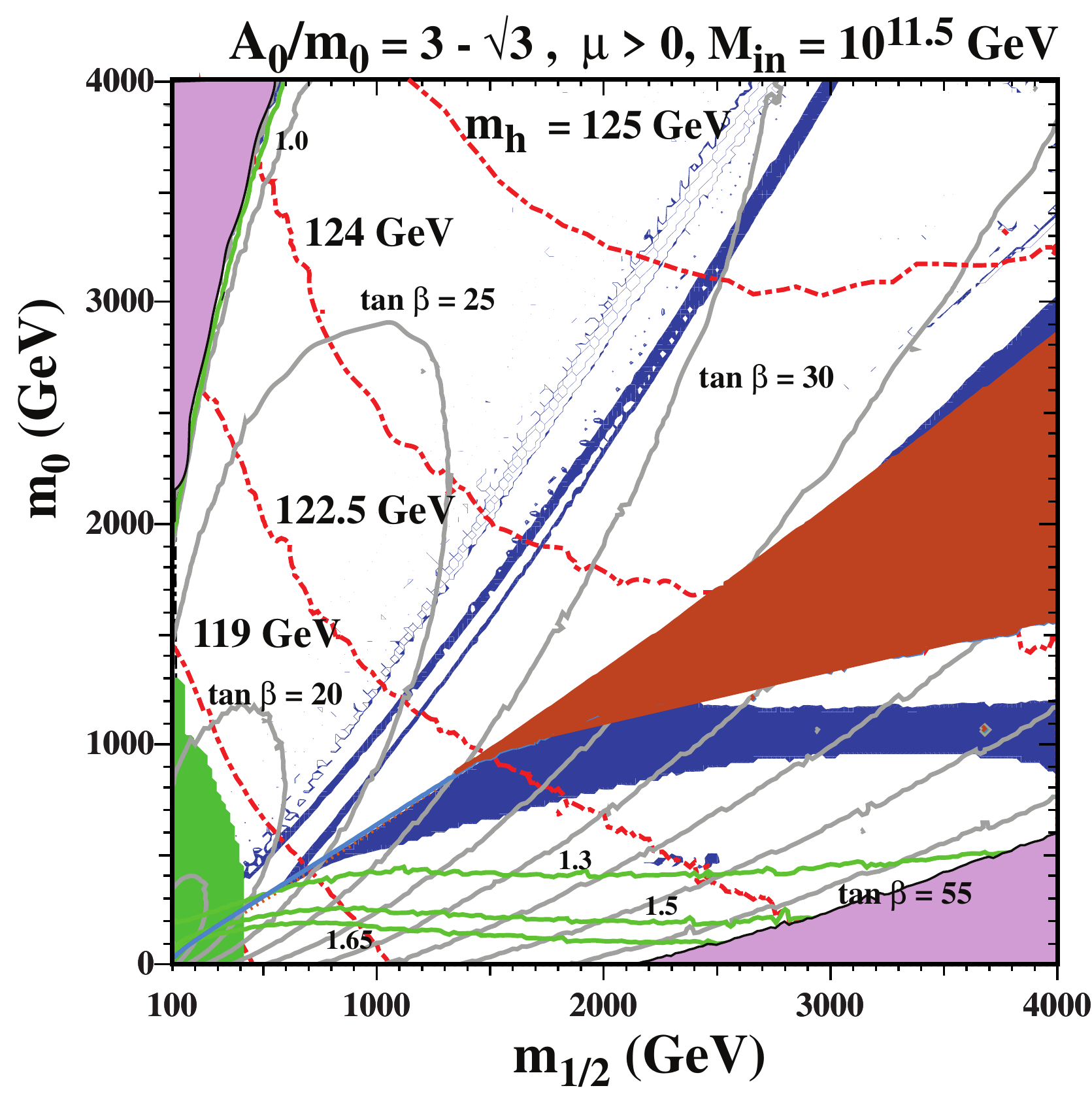}
\hfill
\end{minipage}
\begin{minipage}{8in}
\includegraphics[height=3.3in]{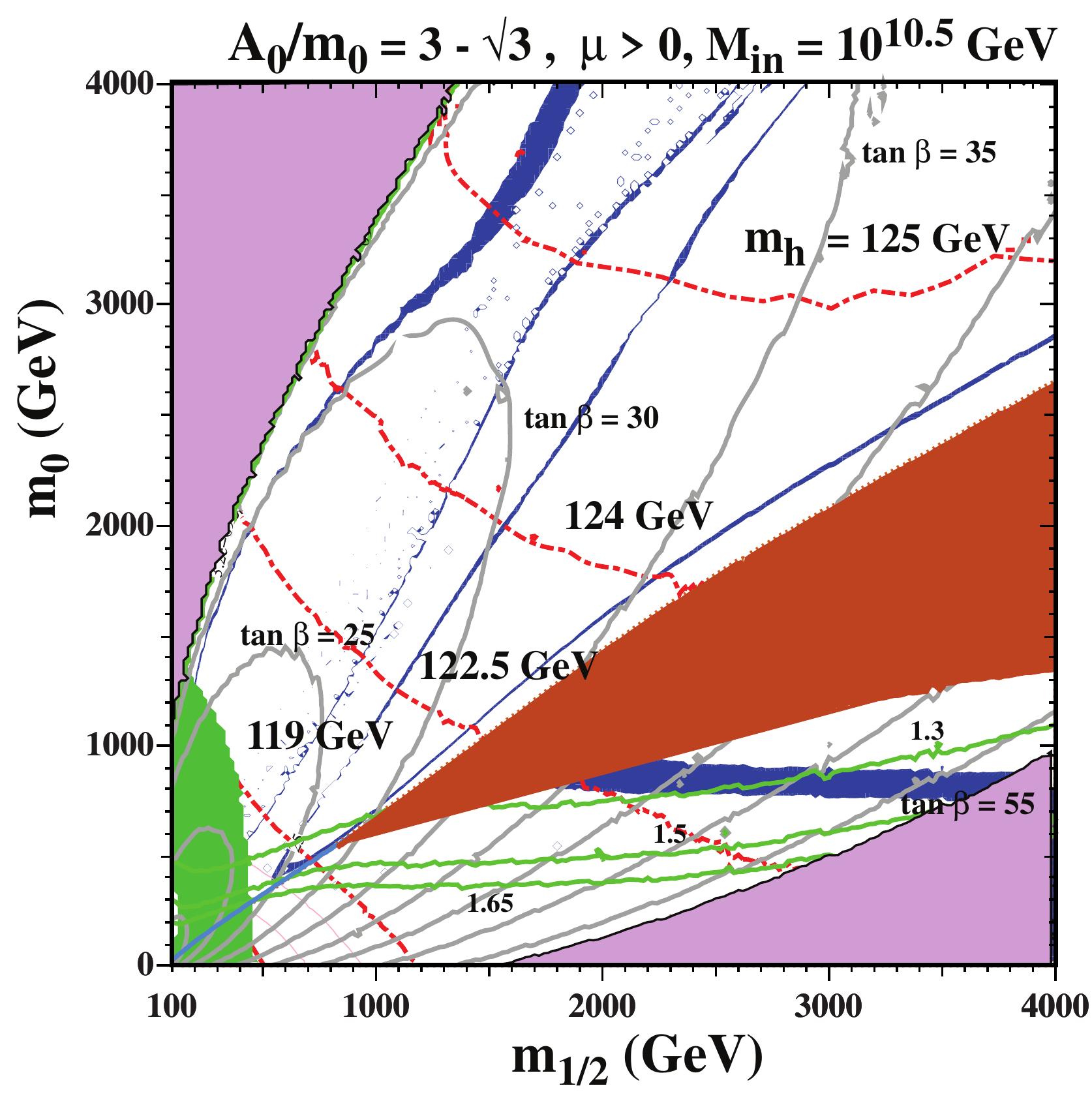}
\hspace*{-0.17in}
\includegraphics[height=3.3in]{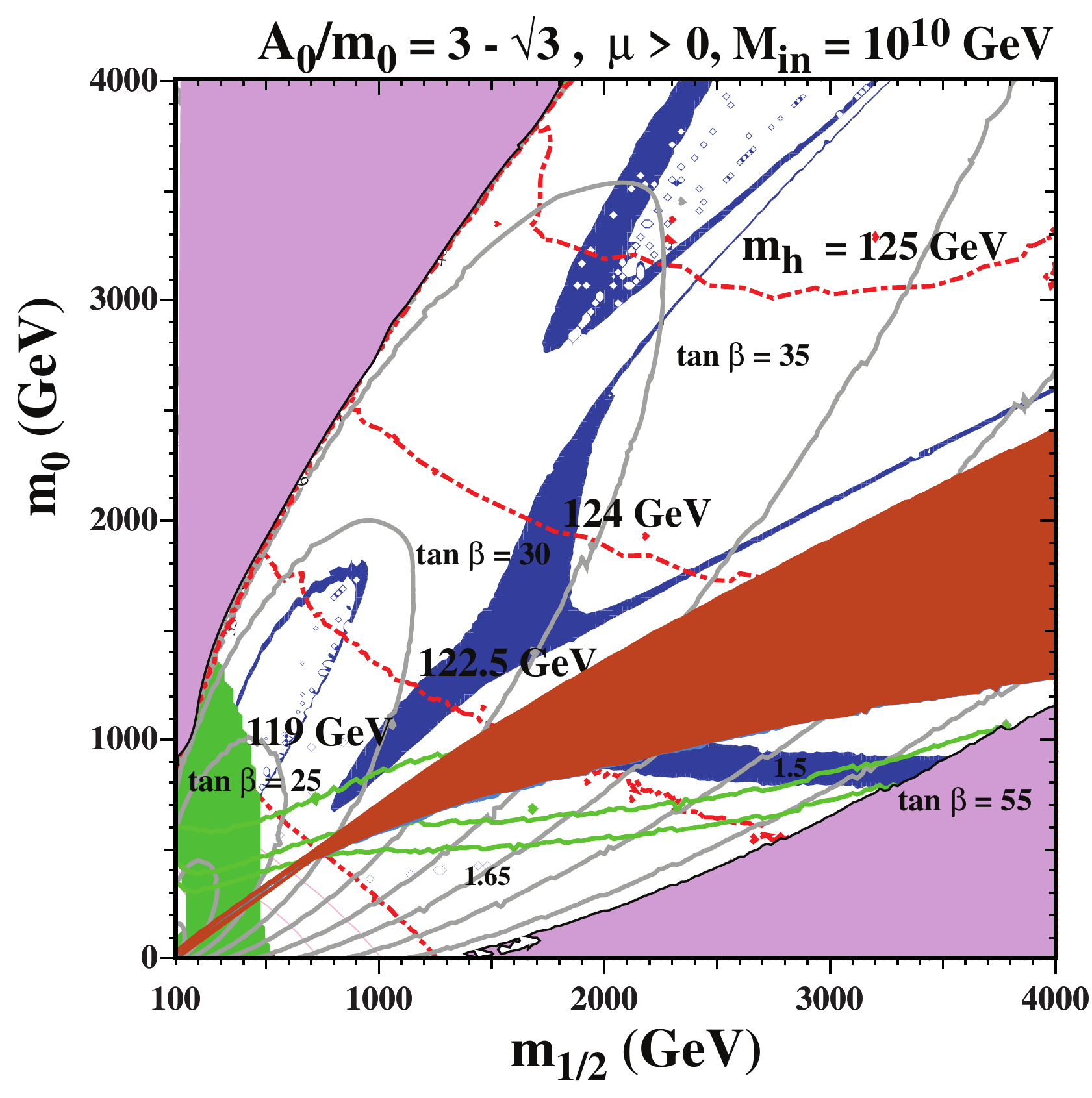}
\hfill
\end{minipage}\caption{
{\it
The $(m_{1/2}, m_0)$ planes for $\mu > 0$ in the sub-GUT Polonyi mSUGRA
model with $A_0 = (3 - \sqrt{3}) \, m_0$ and $M_{in} = 10^{13}$~GeV (upper left), 
$10^{11.5}$~GeV (upper right), $10^{10.5}$~GeV (lower left) and $10^{10}$~GeV (lower right), found
using the latest version of the {\tt SSARD} code~\protect\cite{SSARD}. 
The interpretations of the shadings and contour colours are described in the text.
}} 
\label{fig:Pearlpol} 
\end{figure}

We start with the $(m_{1/2}, m_0)$ planes for the
Polonyi case $A_0 = (3 - \sqrt{3}) \, m_0$ shown in Fig.~\ref{fig:Pearlpol}.
(Results for a GUT-scale Polonyi model were shown in 
the left panel of Figure \ref{fig:msugra}.) The case $M_{in} = 10^{13}$~GeV
is displayed in the upper left panel of  Fig.~\ref{fig:Pearlpol}, where
we see that there is a
forbidden region at large $m_{1/2}$ and $m_0 \sim 2000$~GeV where the ${\tilde \tau_1}$ 
is the LSP. At larger $m_0$ and/or smaller $m_{1/2}$ the lightest neutralino $\chi$ is the LSP, 
and at smaller $m_0$ the gravitino is the LSP. There is a narrow funnel in the $\chi$ LSP
region with suitable cold dark matter density that barely reaches 
$(m_{1/2}, m_0) \sim (2000, 1500)$~GeV, and a `boomerang' region at lower $m_0$,
extending to large $m_{1/2}$, where the density of gravitino LSPs produced in NLSP
decays is suitable for the cold dark matter. The tip of the $\chi$ LSP funnel and the upper
arm of the gravitino boomerang are just barely compatible with $M_h \sim 124$~GeV,
which is compatible with the LHC measurement within uncertainties. We note that all (most of) the
visible part of the plane is compatible with the LHC $B_s \to \mu^+ \mu^-$ (MET)
constraint, including these two $M_h$-compatible regions. In this plane, $\tan \beta$ is 
relatively small except when $m_{1/2}$ is very large.  This accounts for the sufficiently low value
of $B_s \to \mu^+ \mu^-$ relative to the Standard Model (between 1.0 and 1.3 over much of the 
displayed plane). 

Turning now to the upper right panel of Fig.~\ref{fig:Pearlpol} for $M_{in} = 10^{11.5}$~GeV,
and descending from the boundary of the electroweak symmetry breaking region in the
upper left to lower right, across the $\chi$ LSP region, we encounter 
several new dark matter-compatible strips, some of which were present embryonically in 
previous figures. The uppermost in a series of wisps is due to $\chi_3 - \chi^\pm$ coannihilation enhanced
by a near-on-shell t-channel pole. 
Moving across the plane towards larger $m_{1/2}$, several wisps are crossed prior to reaching the 
most prominent diagonal feature, which is the more familiar
$\chi - \chi$ s-channel funnel. 
These wispy features above the $\chi - \chi$ funnel are due to coannihilations of 
neutralinos and charginos enhanced by s-channel $H/A$ poles.
Further down and to the right there is a coannihilation strip
adjacent to the forbidden ${\tilde \tau_1}$ LSP triangle. Continuing below this triangle,
we encounter a near-horizontal band where ${\tilde \tau_1}$ decays yield a suitable
density of gravitino LSPs. In this case, almost all the $(m_{1/2}, m_0)$ plane is
compatible with the LHC MET and $B_s \to \mu^+ \mu^-$ constraints, and large parts 
of the many dark matter-compatible strips in the $\chi$ LSP region are compatible with the
$M_h$ measurement.

\begin{figure}[htb!]
\begin{minipage}{8in}
\includegraphics[height=3.3in]{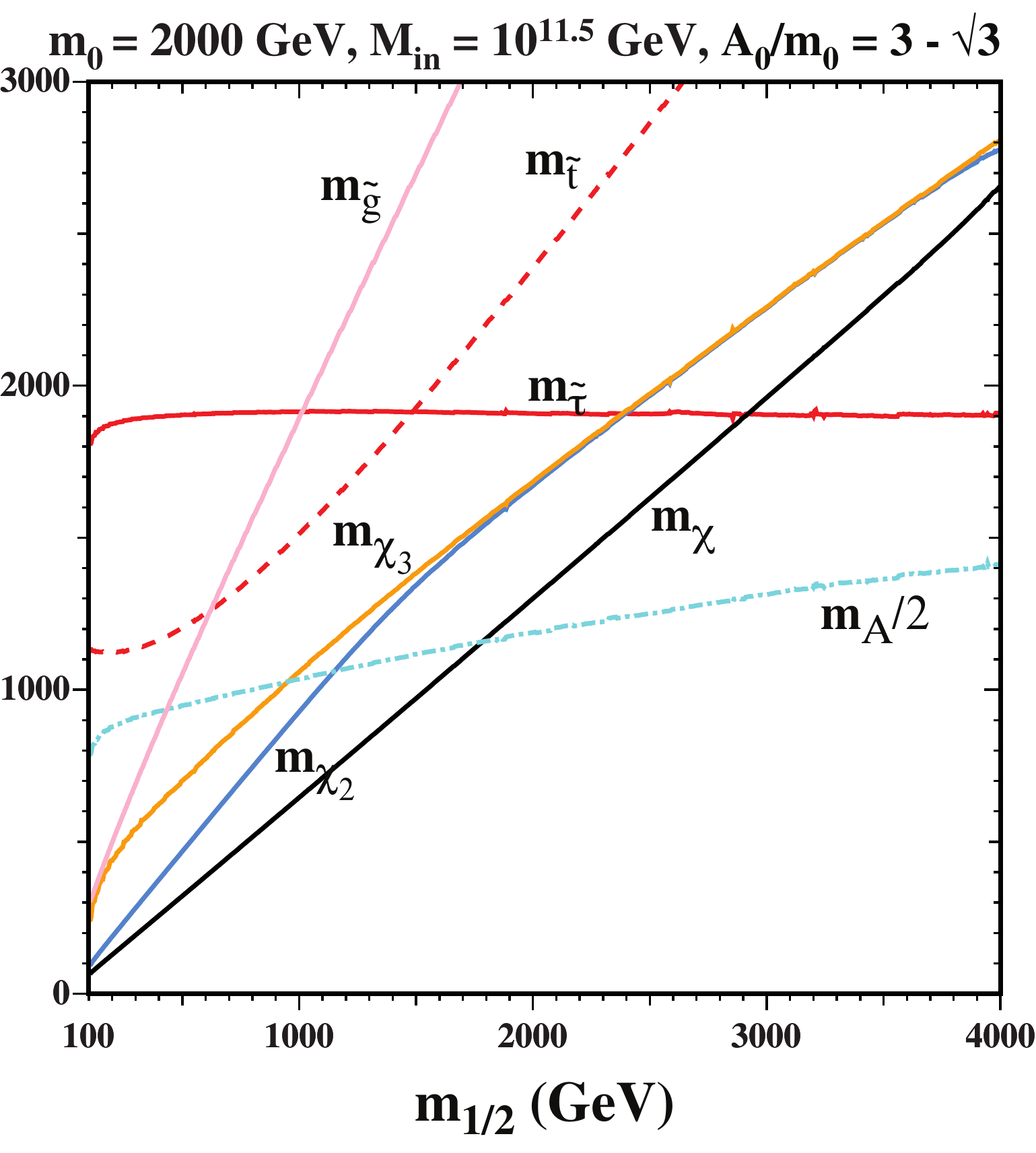}
\hspace*{0.17in}
\includegraphics[height=3.3in]{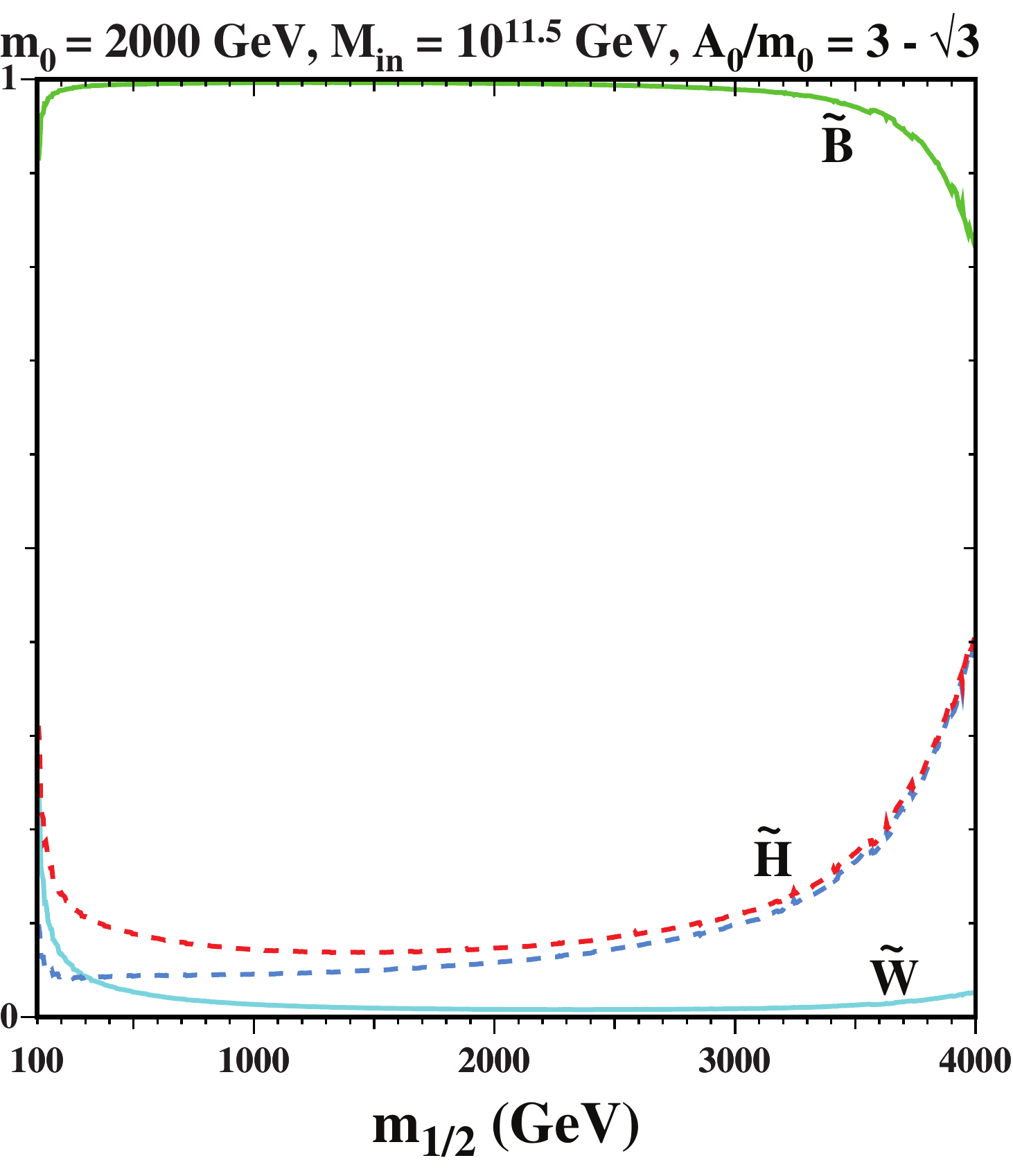}
\hfill
\end{minipage}
\begin{minipage}{8in}
\includegraphics[height=3.3in]{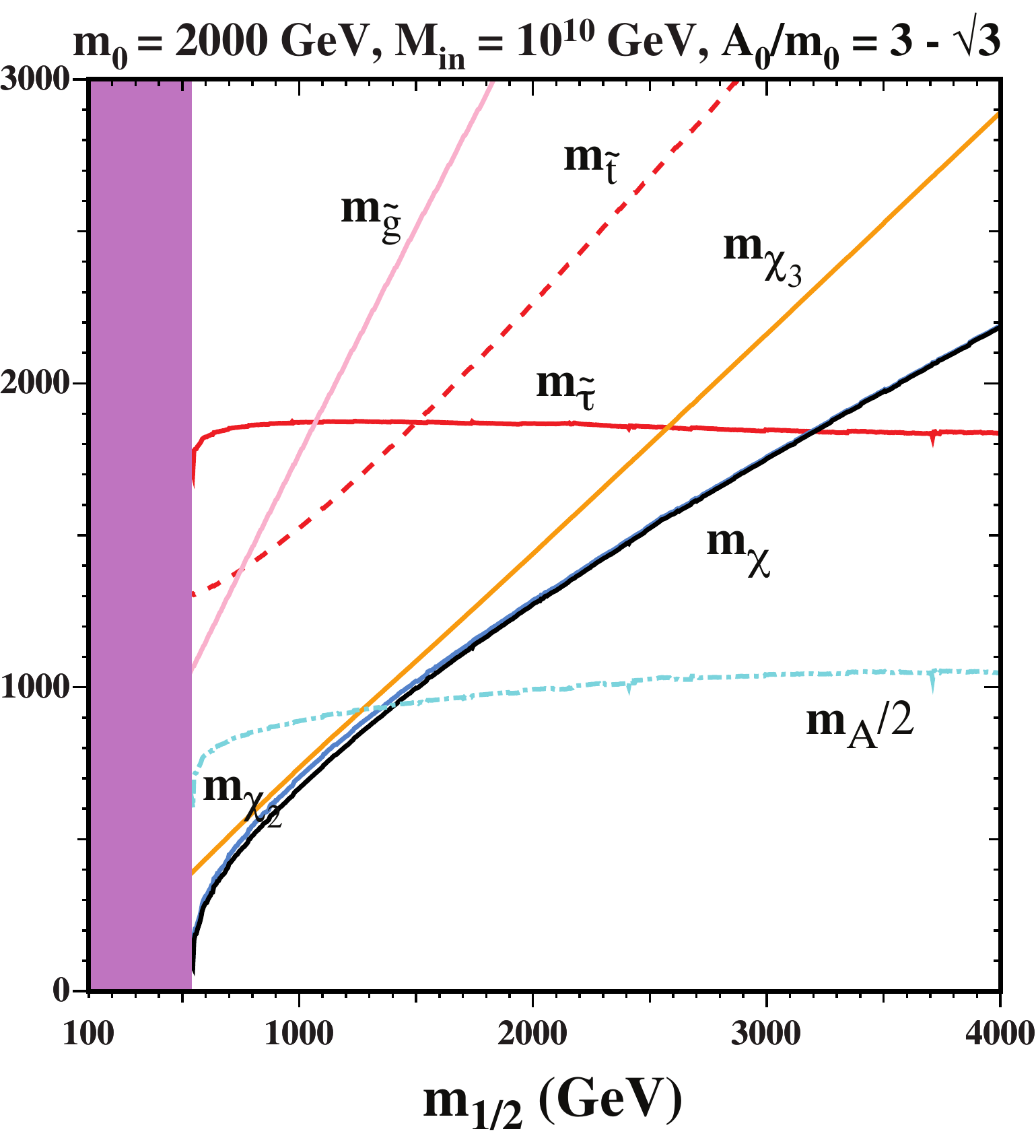}
\hspace*{0.17in}
\includegraphics[height=3.3in]{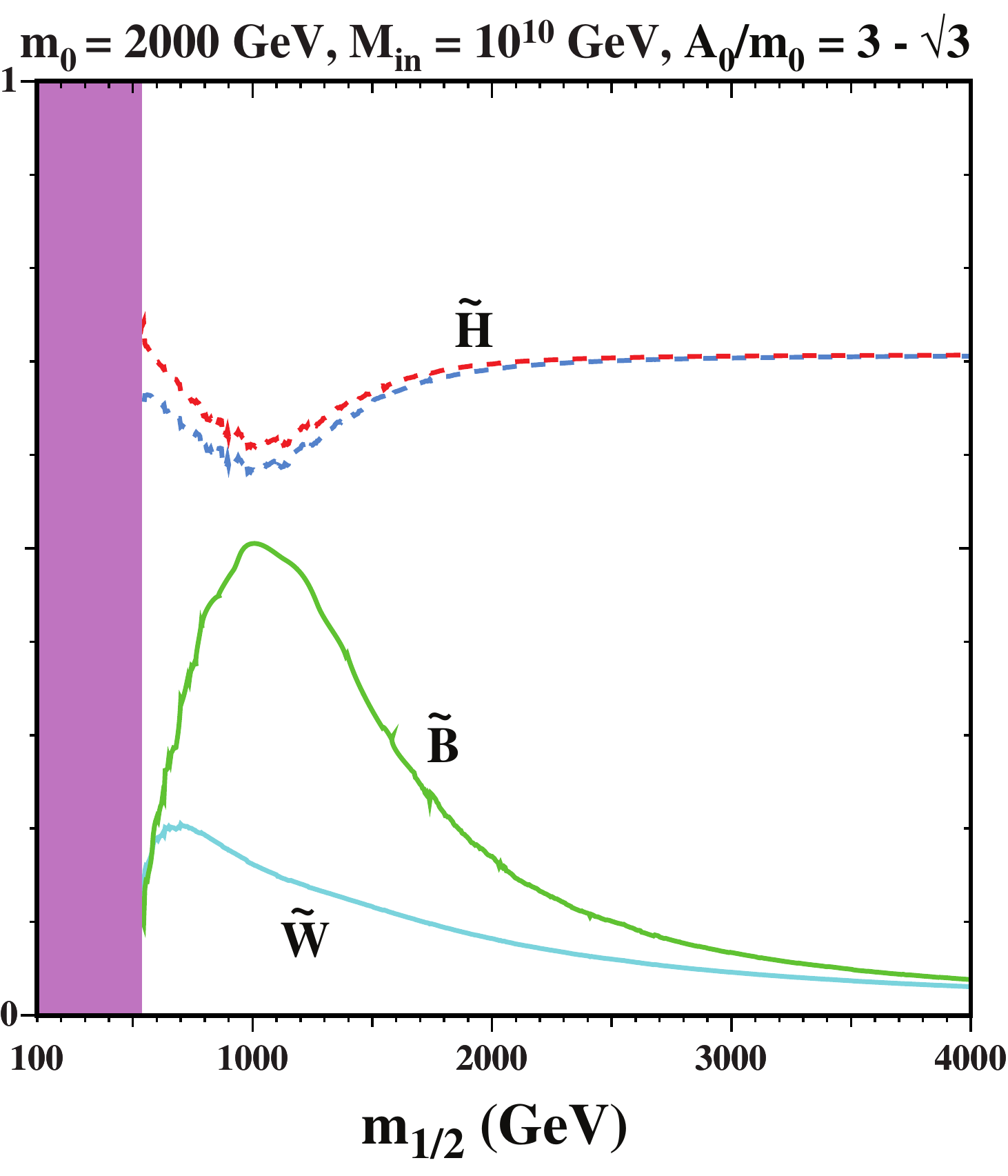}
\hfill
\end{minipage}\caption{
{\it
The sparticle spectrum (left panels) and (right panels) the composition of the
LSP $\chi$ as functions of $m_{1/2}$ in a sub-GUT Polonyi mSUGRA
scenario with $m_0 = 2000$ GeV, for $M_{in} = 10^{11.5}$ GeV (upper panels) and = $10^{10}$ GeV 
(lower panels).}} 
\label{fig:help} 
\end{figure}

To help understand the origin of the multiple avenues for coannihilation,
we show in the upper left panel of Fig.~\ref{fig:help} the sparticle spectrum as a function 
of $m_{1/2}$ for fixed $m_0$ = 2000 GeV for $M_{in}= 10^{11.5}$~GeV.
Here we see clearly the possibility of multiple resonant reactions when $m_A/2$ is equal
to $m_{\chi_3}$ at $m_{1/2} \approx 800$ GeV, $m_{\chi_2}$ at $m_{1/2} \approx 1100$ GeV,
and $m_{\chi}$ at $m_{1/2} \approx 1800$ GeV. In the upper right panel of Fig.~\ref{fig:help}, we see the behaviour of 
the neutralino composition as a function of the universal gaugino mass:
the Higgsino composition is enhanced at large $m_{1/2}$, contributing to 
an increased cross section and hence acceptable relic density past the point where the
stau coannihilation strip would normally end.

Looking back at the $(m_{1/2}, m_0)$ plane for $M_{in} = 10^{10.5}$~GeV shown in the
lower left panel of Fig.~\ref{fig:Pearlpol}, and again descending across the $\chi$
LSP region away from the electroweak symmetry breaking boundary towards the ${\tilde \tau_1}$
LSP region, we again encounter several near-parallel dark matter-compatible strips.
The first is a relatively broad focus-point region where the LSP is more Higgsino-like, 
then there is a coannihilation funnel, followed by the more familiar $\chi - \chi$
rapid-annihilation funnel. Finally, we encounter a Higgsino - ${\tilde \tau_1}$ coannihilation strip.
Below the ${\tilde \tau_1}$ LSP region, there is again a band where ${\tilde \tau_1}$
NLSP decays to gravitinos yield a suitable relic dark matter density. As in the previous case,
most of the $(m_{1/2}, m_0)$ plane is again compatible with the LHC MET and 
$B_s \to \mu^+ \mu^-$ constraints due to the relatively large mass scales and/or small value of 
$\tan \beta$.  The upper parts ($m_0 \ga 2$ TeV) of the dark matter-compatible strips 
in the $\chi$ LSP region are compatible with the $M_h$ measurement.

Finally, we consider the $(m_{1/2}, m_0)$ plane for $M_{in} = 10^{10}$~GeV shown in the
lower right panel of Fig.~\ref{fig:Pearlpol}. At large $m_0$, we see that the broad focus-point
strip is joined to one side of the standard $\chi - \chi$ funnel. The other side of this funnel
is connected to a Higgsino coannihilation strip hovering above the ${\tilde \tau_1}$ LSP
boundary. At lower $m_0$ and $m_{1/2}$, there is another loop connecting the lower part
of the focus-point strip with the $\chi - \chi$ funnel, and at low $m_0$ but larger $m_{1/2}$
there is again a band where gravitinos could provide the dark matter. As before, most of the
$(m_{1/2}, m_0)$ plane is compatible with the LHC MET and $B_s \to \mu^+ \mu^-$
constraints, and there are large stretches of the dark matter strips at large $m_{1/2}$
and $m_0$ that are compatible with the LHC Higgs measurement.

For this case also, we display the sparticle masses and neutralino composition as
functions of $m_{1/2}$ for $m_0 = 2000$ GeV in the lower panels of Figure \ref{fig:help}.
Here, we clearly see the congruence of masses at $m_{1/2} \approx 1000$ GeV 
all contributing to a resulting low relic density.  In this case, we also see that, for this value of $m_0$,
the neutralino is predominantly a Higgsino, but in the funnel centre, there is an enhanced
bino component leading to a very mixed LSP which also leads in general to an enhanced cross section.

The above discussion was for sub-GUT Polonyi mSUGRA models, and we now
discuss briefly what happens in sub-GUT mSUGRA models with values
of $A_0 \ne (3 - \sqrt{3}) \, m_0$. First, we recall that in mSUGRA models the value
of $\tan \beta$ is fixed by the electroweak conditions as a function of the other,
independent model parameters, as shown by the grey contours in the different
panels of Fig.~\ref{fig:Pearlpol}. As shown there, in the Polonyi case typical values are
$\tan \beta \sim 20$ to 30, which is why the $B_s \to \mu^+ \mu^-$ has relatively
little impact - though this does increase slightly at smaller $M_{in}$ where larger
values of $\tan \beta$ are found. On the other hand, we have already seen
in the right panel of Fig. \ref{fig:msugra} that the derived values of $\tan \beta$ are as large as $\sim 40$
for $M_{in} = M_{GUT}$. 

\begin{figure}[htb!]
\begin{minipage}{8in}
\includegraphics[height=3.3in]{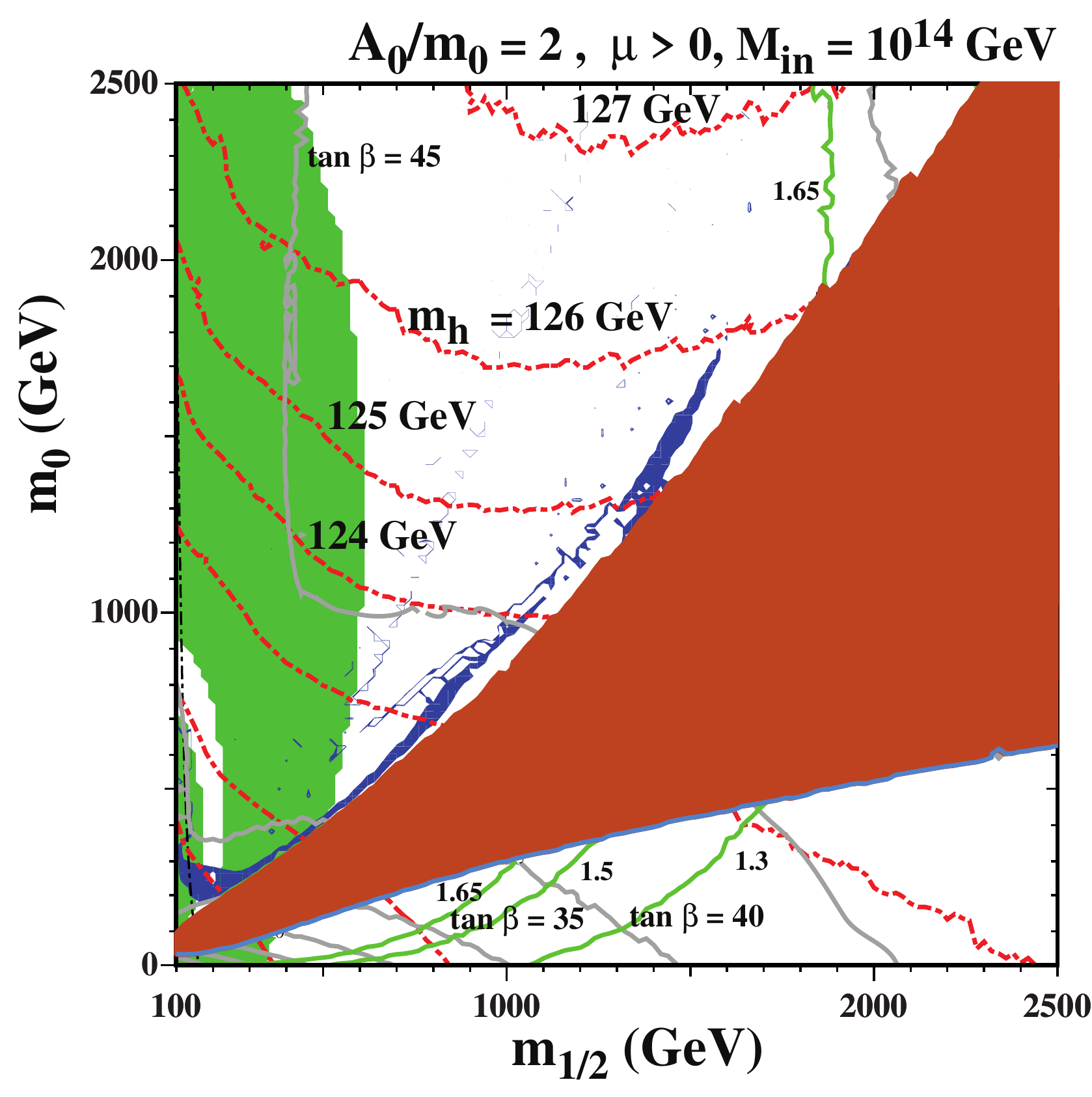}
\hspace*{-0.17in}
\includegraphics[height=3.3in]{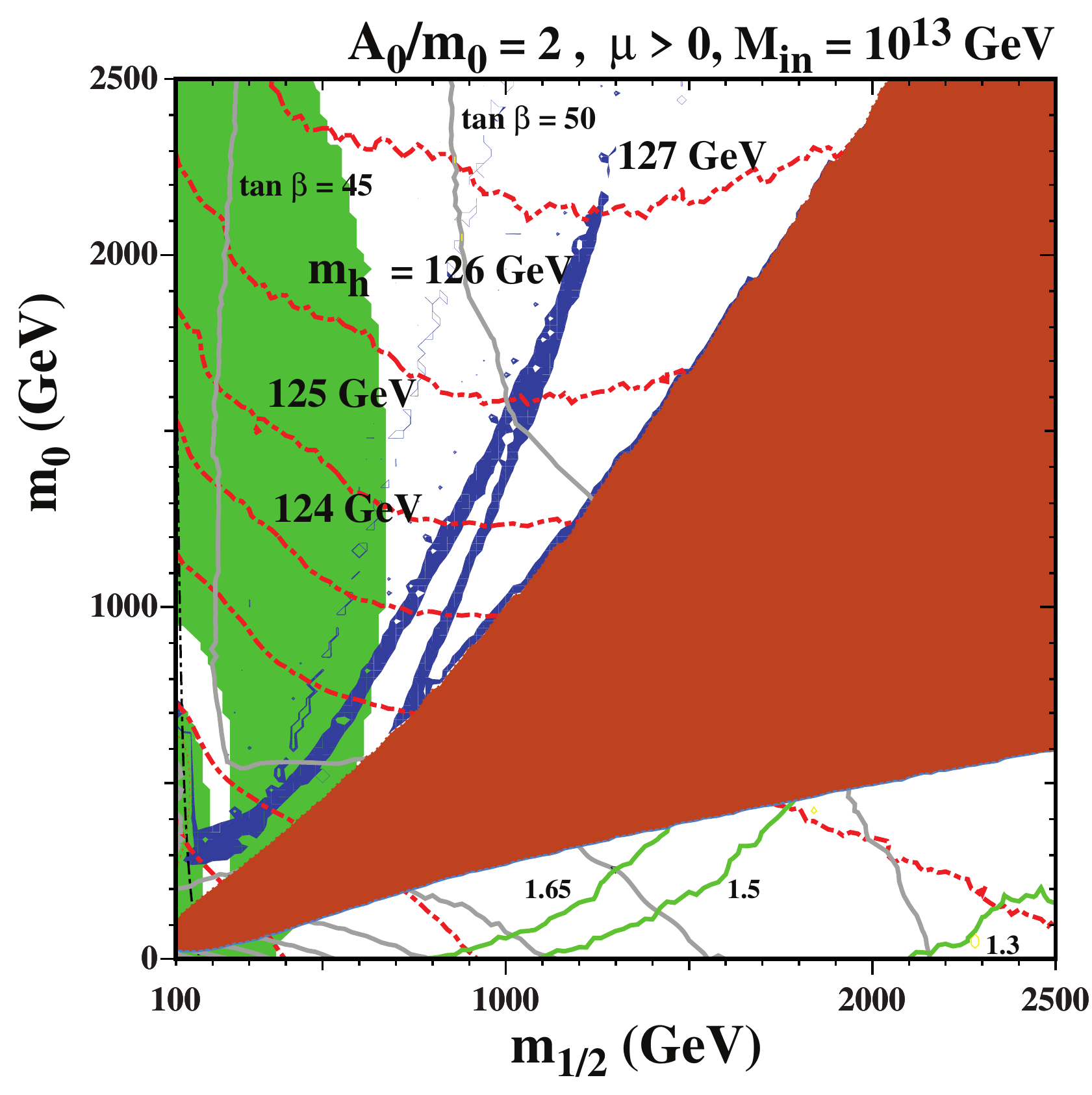}
\hfill
\end{minipage}
\caption{
{\it
The sub-GUT mSUGRA $(m_{1/2}, m_0)$ planes for $A_0 = 2 \, m_0$ and
$M_{in} = 10^{14}$~GeV (left), $M_{in} = 10^{13}$~GeV (right).
The interpretations of the shading and contour colours are described in the text.
}}
\label{fig:Pearl2} 
\end{figure}

In sub-GUT mSUGRA models with $A_0 = 2 \, m_0$, as shown in Fig.~\ref{fig:Pearl2}, we see that
$\tan \beta$ increases to $\sim 45$ to 50 for $M_{in} = 10^{14}$ and $10^{13}$~GeV. 
This has two consequences worth noting: (i) the impact of the $B_s \to \mu^+ \mu^-$
constraint is increased to such an extent, compared to the Polonyi case, that no
$A_0 = 2 \, m_0$ sub-GUT mSUGRA models are valid for either value of $M_{in}$, and (ii)
for $M_{in} < 10^{13}$~GeV, the increasing value of $\tan \beta$ induces non-perturbative
instabilities in the renormalization-group equations over much of the $(m_{1/2}, m_0)$
plane. For these reasons, sub-GUT mSUGRA models with $A_0$ much larger than the
Polonyi value are disfavoured.

In the case of sub-GUT mSUGRA models with $A_0$ less than the Polonyi value,
we recall (see also the left panel of Fig.~\ref{fig:msugra}), that $M_h$ is much smaller
than the LHC measurement if $A_0 = 0$ and $M_{in} = M_{GUT}$. This problem
persists for $M_{in} < M_{GUT}$, so we do not discuss further this possibility.

The key messages from this study are that (i) many (co)annihilation processes that
are relatively unimportant in the CMSSM can become important in sub-GUT models,
as the sparticle spectrum becomes more degenerate and Boltzmann suppression factors
therefore become less significant. As a result, (ii) the cold dark matter density may fall
into the favoured range almost anywhere in the interior of the $(m_{1/2}, m_0)$
plane away from the ${\tilde \tau_1}$ LSP and electroweak symmetry breaking boundaries,
depending on the precise value of $M_{in}$ and $A_0$. Next, (iii) many of the
regions allowed by the dark matter constraint are compatible with the current LHC
constraints, including the $M_h$ measurement. Finally and intriguingly, (iv) in the specific cases of
sub-GUT mSUGRA models, values of $A_0$ very different from the Polonyi value
$A_0 = (3 - \sqrt{3}) \, m_0$ are disfavoured.

\section{Summary}

It is well known that only very small regions of the CMSSM
parameter space are consistent with all the theoretical, phenomenological, 
experimental, astrophysical and cosmological constraints, notably
including the preferred range of the cold dark matter density.
We have explored in this paper the extent to which some generalizations
of the CMSSM may wriggle out of the straitjacket imposed, in particular, by the LHC 
constraint on missing-energy (MET) events, by the LHCb measurement of the
branching ratio for $B_s \to \mu^+ \mu^-$ and the LHC measurement
of the mass of the Higgs boson. The principal classes of models studied
are those with non-universal supersymmetry-breaking contributions to the
Higgs masses (NUHM1,2) and models in which universality is imposed on the
supersymmetry-breaking parameters at some input scale $M_{in} < M_{GUT}$
(sub-GUT CMSSM and mSUGRA).

It is not difficult to find regions of the NUHM1,2 parameter spaces where the
cosmological cold dark matter density falls within the preferred range, even
if the sparticle masses are relatively large, as required by the LHC MET and
$M_h$ constraints. For example, this may happen in a transition region where
the $\chi$ LSP has a relatively large Higgsino component. In the CMSSM, this
could happen only in the focus-point region, a possibility now disfavoured
by the XENON100 upper limit on cold dark matter scattering, but in the NUHM1 the $\mu$
parameter may be regarded as free and the transition region may appear
in other parts of the $(m_{1/2}, m_0)$ plane where the LHC MET and $M_h$
constraints are respected and BR$(B_s \to \mu^+ \mu^-)$ is close to the
Standard Model value.

Sub-GUT models offer other possibilities for reconciling the LHC and other
constraints. The compression of the spectrum as $M_{in}$ decreases implies
that more coannihilation processes may become important, suppressing the
relic density below the range expected in the CMSSM, and enabling it to
be compatible with the cosmological constraint also at relatively large values
of $m_{1/2}$ and $m_0$. Also, in sub-GUT models we find a proliferation
of strips where the dark matter density falls within the preferred range.

The NUHM1,2 and sub-GUT models are examples to illustrate that there
is no general conflict between the LHC constraints and supersymmetry,
even within the MSSM framework. It suffices to relax the constraint of
universality of the soft supersymmetry-breaking parameters at the GUT
scale, as we have demonstrated in a few types of scenarios. However,
it is worth noting that the surviving models in these particular scenarios tend to
have quite large sparticle masses, and do not provide a supersymmetric
resolution of the $g_\mu - 2$ discrepancy.

\section*{Acknowledgments}
The work of
J.E. and F.L. was supported in part by the London Centre for Terauniverse
Studies (LCTS), using funding from the European Research Council via
the Advanced Investigator Grant 267352: this also supported
visits by K.A.O.  to the CERN TH Division, which he thanks
for its hospitality. The work of F.L. and K.A.O. was supported in part
by DOE grant DE--FG02--94ER--40823 at the University of Minnesota, and 
the work of F.L. was also supported in part by a Doctoral Dissertation Fellowship at the University of Minnesota. 
P.S.~gratefully acknowledges support and resources from the Center for High Performance Computing at the University of Utah.

\end{document}